\documentclass[pre,twocolumn,aps,floatfix,superscriptaddress]{revtex4}

\usepackage[english]{babel}
\usepackage{amssymb,amsfonts,amsmath}
\usepackage[pdftex]{graphicx}
\usepackage{latexsym}
\usepackage{multirow}
\usepackage{color}

\begin{document}

\title{Rethinking the logistic approach for population dynamics of mutualistic interactions}

\author{Javier Garc\'{\i}a-Algarra}\affiliation{Complex System Group, Universidad Polit\'ecnica de Madrid, 20040 Madrid, Spain}

\author{Javier Galeano}\affiliation{Complex System Group, Universidad Polit\'ecnica de Madrid, 20040 Madrid, Spain}\affiliation{Dep. Ciencia y Tecnolog\'{\i}a Aplicadas a la I.T. Agr\'{\i}cola, E.U.I.T. Agr\'{\i}cola, 
Universidad Polit\'ecnica de Madrid, 20040 Madrid, Spain}

\author{Juan Manuel Pastor}\affiliation{Complex System Group, Universidad Polit\'ecnica de Madrid, 20040 Madrid, Spain}\affiliation{Dep. Ciencia y Tecnolog\'{\i}a Aplicadas a la I.T. Agr\'{\i}cola, E.U.I.T. Agr\'{\i}cola, 
Universidad Polit\'ecnica de Madrid, 20040 Madrid, Spain}

\author{Jos\'e Mar\'{\i}a Iriondo}\affiliation{\'Area de Biodiversidad y Conservaci\'on, Dept. Biolog\'{\i}a y Geolog\'{\i}a, Universidad Rey Juan Carlos, 28933 M\'ostoles, Spain}

\author{Jos\'e J. Ramasco}\affiliation{Instituto de F\'{\i}sica Interdisciplinar y Sistemas Complejos IFISC (CSIC-UIB), Campus UIB,\\ 07122 Palma de Mallorca, Spain}

\widetext

\begin{abstract}
Mutualistic communities have an internal structure that makes them resilient to external perturbations. Late research has focused on their stability and the topology of the relations between the different organisms to explain the reasons of the system robustness. Much less attention has been invested in analyzing the systems dynamics. The main population models in use are modifications of the \emph{r - K} formulation of logistic equation with additional terms to account for the benefits produced by the interspecific interactions. These models have shortcomings as the so called \emph{r - K} formulation diverges under some conditions. In this work, we introduce a model for population dynamics under mutualism that preserves the original logistic formulation. It is mathematically simpler than the widely used type II models, although it shows similar complexity in terms of fixed points and stability of the dynamics. We perform an analytical stability analysis and numerical simulations to study the model behavior in general interaction scenarios including tests of the resilience of its dynamics under external perturbations. Despite its simplicity, our results indicate that the model dynamics shows an important richness that can be used to gain further insights in the dynamics of mutualistic communities. 
\end{abstract}

\maketitle

\section{Introduction}
\label{intro}

Despite its long history, there are still several open issues in the research of ecological population dynamics. Some of these questions were highlighted in the 125th anniversary issue of the journal {\em Science} \citep{Kennedy05,Pennisi05,Stokstad05}. For example, aspects such as the mechanisms determining species diversity in an ecosystem are under a very active scrutiny by an interdisciplinary scientific community \citep{williams00,dunne02,olensen07,allesina08,bascompte09,saavedra09,bastolla09,fortuna2010nestedness,encinas12}.  
Quantitative population dynamics goes back to $1202$ when Leonardo Fibonacci, in his {\em Liber Abaci}, described the famous series that follows the growth of rabbit population \citep{Sigler02}. Classical population theory began, however, in $1798$ with Robert Malthus' {\em An Essay on the Principle of Population} \citep{Malthus98}. Malthus argued that population growth is the result of the difference between births and deaths, and that these magnitudes are proportional to the current population. Mathematically, this translates in the differential equation:
\begin{equation}
\frac{dN}{dt}=r_0\, N ,
\label{eq:malthus}
\end{equation}
where $N$ is the population size, $r_0$ is the {\em intrinsic rate} of growth of the population and equals the difference between the rates of birth and death (assuming no migrations). 

The Malthusian model predicts an exponential variation of the population, which if $r_0 > 0$ translates into an unbounded growth. In this model, $r_0$ remains constant along the process ignoring thus limiting factors on the population such as the lack of nutrients or space. In $1838$ Verhulst introduced an additional term, proposing the so-called \emph{logistic} equation (see \cite{Verhulst1845}). The growth rate  must decrease as $N$ increases to limit population growth and the simplest way to achieve this is by making $r_0$ a linear function of $N$: $ r_0 = r - \alpha N$, where $r$ is the intrinsic growth rate and $\alpha$ a positive (friction) coefficient that is interpreted as the intraspecific competition. This approach leads to the $r-\alpha$ model:
\begin{equation}
\frac{dN}{dt}=r \, N \,  - \alpha  \, N^2 .
\label{eq:primitiveverhulst}
\end{equation}
The term with $\alpha$ acts as a biological \emph{brake} leading the system to a point of equilibrium for the dynamics with a population value approaching $ K = r / \alpha$, usually called the \emph{carrying capacity} of the system.

The logistic equation is best known in the form that Raymond Pearl introduced in $1930$ (see \cite{mallet2012struggle} for an excellent historical review). In this formulation, the carrying capacity appears explicitly, and so it is known as $r-K$:

\begin{equation}
\frac{dN}{dt}=r \, N \, \left(1-\frac{N}{K}\right) .
\label{pearl}
\end{equation}
The solution of this equation is a sigmoid curve that asymptotically tends to $K$. This formulation has some major mathematical drawbacks \citep{kuno1991some,gabriel2005paradoxes}. The most important is that it is not valid when the initial population is higher than the carrying capacity and $r$ is negative. Under those conditions, it predicts an unbounded population growth. This issue was noted by Richard Levins, and consequently is called the Levins' paradox \citep{gabriel2005paradoxes}. It is important to stress that all mutualistic models derived from Pearl's formula inherit its limitations in this sense.

These seminal models of population dynamics did not take into account interactions between species. When several species co-occur in an community there can be a rich set of relationships among them that can be represented as a complex interaction network. In $1926$, Vito Volterra proposed a two-species model to explain the behavior of some fisheries in the Adriatic sea \citep{Volterra26}. Volterra's equations describe prey $N(t)$ and predator populations $P(t)$ in the following way: 
\begin{align}
\displaystyle &\frac{dN}{dt}=N\, \left(a-b \,P\right), \nonumber\\
\displaystyle &\frac{dP}{dt}=P\, \left(c\, N-d\right) , 
\label{myeq1}
\end{align}
where $a$, $b$, $c$, and $d$ are positive constants. In the Lotka-Volterra model, as it is known today, the prey population growth is limited by the predator population, while the latter benefits from the prey and is bounded by its own growth. This pair of equations has an oscillatory solution that in the presence of further species can even become chaotic.

While prey-predator and competition interactions have been extensively studied, mutualistic interactions, which are beneficial for all the species involved, have received a lower level of attention. Interestingly, back in the XIX century, Charles Darwin had already noticed the importance of a mutualistic interaction between orchids and their pollinators \citep{Darwin62}. Actually, the relations between plants and their pollinators and seed dispersers are the paradigmatic examples of mutualism. In this context, \citet{Ehr64} alluded to the importance of plant-animal interactions in the generation of Earth's biodiversity. The simplest mutualistic model without {\it 'an orgy of mutual benefaction'} was proposed by \citet{may1981models}. Each of May's equations for two species is a logistic model with an extra term accounting for the mutualistic benefit. It is the same idea as in the Lotka-Volterra model but interactions between species always add to the resulting population. May's equations for two species can be written as
\begin{align}
\frac{dN_1}{dt}=r_1 \,N_1\,\left(1-\frac{N_1}{K_1}\right)+r_1\, N_1\,\beta_{12}\, \frac{N_2}{K_1} , \nonumber \\ 
\frac{dN_2}{dt}=r_2\, N_2\, \left(1-\frac{N_2}{K_2}\right)+r_2\, N_2\, \beta_{21} \, \frac{N_1}{K_2} , 
\label{myeq2}
\end{align}
where $N_1(N_2)$ is the population of the species $1(2)$; $r_1 \, (r_2)$ is the intrinsic growth rate of population $1\, (2)$ and $K_1\, (K_2)$ the carrying capacity. This is the maximum population that the environment can sustain indefinitely, given food, habitat, water and other supplies available in the environment. Finally, $\beta_{12}\,(\beta_{21})$ is the coefficient that embodies the benefit for population $1\,(2)$ of each interaction with population $2\,(1)$. May's model major drawback is that it also leads to unbounded growth. This model has been, anyhow, an inspiration for subsequent mutualist models that incorporate terms to solve this problem.

Different strategies to avoid the unlimited growth have been adopted. \citet{Wright89} proposed a two-species model with saturation as a result of restrictions on handling time, $T_H$, which corresponds to the time needed to process resources (food) produced by the mutualistic interaction. The mutualistic term can be included as a type II functional response 
\begin{align}
\frac{dN_1}{dt}=r_1\, N_1\, - \alpha_1 \, N_1^2+ \frac{a\, b\, N_1\,N_2}{1+ a\, N_2\,T_H} \nonumber,\\
\frac{dN_2}{dt}=r_2\, N_2\, - \alpha_2 N_2^2 + \frac{a\,b\,N_1\,N_2}{1+a\, N_1\, T_H} ,
\label{eq_typeII}
\end{align}
where $a$ is the effective search rate and $b$ is a coefficient that accounts for the rate of encounters between individuals of  species $1$ and $2$. Wright analyzes two possible behaviors of mutualism: \emph{facultative} and \emph{obligatory}. In the facultative case, $r_{1,2}$ are positive, {\em i.e.}, mutualism increases the population but it is not indispensable to species subsistence. If $r_{1,2}$ are negative mutualism is mandatory to the  species survival. This model has different dynamics depending on the parameter values, but for a very limited region of parameters shows three fixed points. One stable at both species extinction, another also stable at large population values and a saddle point separating both basins of attractions. Using a mutualistic model with a type II functional, \citet{Bastolla05,bastolla09} show the importance of the structure of the interaction network to minimize competition between species and to increase biodiversity. The type II models are, however, hard to treat analytically due to the fractional nature of the mutualistic term. Other recent alternatives have been proposed as, for instance, that of \citet{johnson13}. Still, these works go in the direction of adding extra features to the type II functional rendering more difficult an eventual analytical treatment. 

Recently, the research in this area has focused on system stability, looking for an explanation of the resilience of these communities in the interaction networks \citep{saavedra09,bastolla09,thebault2010stability, fortuna2010nestedness, staniczenko2013ghost}. The dynamics is, however, as important since changes in the parameters that govern the equations induced by external factors can lead the systems to behave differently and to modify their resilience to perturbations in the population levels. Here, we revisit the basic model describing the population dynamics and propose a set of new equations that combines simplicity in its formulation with the richness of dynamical behaviors of the type II models. 

Once introduced the classical population dynamics equations and the review of mutualistic models, the paper is organized as follows. In Section 2, we propose a modified logistic model for mutualism, along with its stability analysis in Section 3. Numerical simulations of our model studying resilience to external perturbation or to changes in the interaction networks are presented in Section 4. The work is then closed in Section 5 with the conclusions. More technical aspects are considered in Appendices A, B and C with details on the stability analysis and numerical treatment of the equations in stochastic form, as well as the tables with the parameters used for the simulations.

\section{A logistic equation for population dynamics with mutualistic interactions} 
\label{model}

Our basic hypothesis is that mutualism contributes to a variation in the species intrinsic growth rate. This assumption is based on empirical observations in which the growth rate of populations (or the fertility) correlates with the availability of resources (see, for instance, \cite{stenseth98,krebs02,rueness03,tyler08,jones08}). In our context, the resources are provided by the mutualistic interactions. The simplest way to express mathematically this idea is by expanding the intrinsic growth rate $r$ in terms of the populations with which the mutualistic interactions occur. To be more specific, let us assume that the community is composed of $n_a$ animal species with populations $\{N_i^a\}$ and $n_p$ plant species with populations $\{N_j^p\}$. The rate of mutualistic interactions between a species $i$ and another $j$ is given by $b_{ij}$, which can be seen as elements of a matrix encoding the mutualistic interaction network. Note that the matrix is not necessarily symmetric if the benefit of the interaction is different for the two species involved. Considering a generic animal species $i$, its growth rate can then be written as
\begin{align}
r_{i} = r_{i}^{0} + \sum_{k=1}^{n_{p}} b_{ik}\, N^{p}_k,
\label{eq:expr}
\end{align}
where $r_{i}^{0}$ is the initial vegetative growth rate. To avoid unrealistic divergence in the population levels, the effect of mutualism must saturate at certain point. Following Velhurst's idea for the logistic equation, this implies that the friction term, $\alpha_i$, must also depend on the mutualistic interactions. In order to keep the model simple, we assume that the effect of the mutualism on $\alpha$ is proportional to the benefit. This means that  
\begin{align}
\alpha_i = \alpha_{i}^{0}+ c_{i} \sum_{k=1}^{n_{p}} b_{ik}\, N^{p}_k ,
\stepcounter{equation}\tag{\theequation}
\label{eq:alphavariable}
\end{align}
where $c_{i}$ is a proportionality constant. The expansions for the plants are similar but with the sums running over the animal species instead of on the plants. The expansions of $r$ and $\alpha$ could have been taken to higher orders in $N^{p}_k$. However, the linear version of the model should be enough to capture the qualitative features of the population dynamics as long as the higher order terms contribute in a similar way to $\alpha$ and $r$ (with the same sign). 

For the sake of simplicity in the notation whenever there is no possible confusion the zeros will be dropped from $\alpha_{i}^{0}$ and $r_{i}^{0}$. Under these assumptions, the system dynamics is described by the following set of differential equations:
\begin{align}
\frac{1}{N^{a}_{i}}\frac{dN^{a}_{i}}{dt} = r_{i}+ \sum_{k=1}^{n_{p}} b_{ik}\, N^{p}_k - \left( \alpha_{i}+ c_{i} \sum_{k=1}^{n_{p}} b_{ik}\, N^{p}_k \right) N^{a}_{i} \nonumber\\
\frac{1}{N^{p}_{j}}\frac{dN^{p}_{j}}{dt} = r_{j}+ \sum_{\ell=1}^{n_{a}} b_{j\ell}\, N^{a}_\ell - \left( \alpha_{j}+ c_{j} \sum_{\ell=1}^{n_{a}} b_{j\ell}\, N^{a}_\ell \right) N^{p}_{j}
\stepcounter{equation}\tag{\theequation}\label{eq:modeloralphaconmut}
\end{align}
The terms on the right-hand side of these equations can be interpreted as a \emph{effective growth rates}. Since we will use this concept later, it is important to define it explicitly. The effective growth rate of an animal species $i$ is defined as
\begin{equation}
r_{ef,i} = r_{i}+ \sum_{k=1}^{n_{p}} b_{ik}\, N^{p}_k - \left( \alpha_{i}+ c_{i} \sum_{k=1}^{n_{p}} b_{ik}\, N^{p}_k \right) N^{a}_{i} .
\label{eq:effrate}
\end{equation}
The plants effective growth rates are defined equivalently but substituting $a$ by $p$. The \emph{carrying capacities} of the system are given by the non-zero fixed points of Eqs. (\ref{eq:modeloralphaconmut}). It is easy to see that in the absence of mutualism $K_i = r_i/\alpha_i$ for species $i$, as in the original logistic equations. On the other hand, under the presence of very strong mutualism $K_i$ tends to $1/c_{i}$. The role of the proportionality constant $c_i$ is thus to establish a maximum population for the species $i$ in the strong interaction limit $c_{i} \sum_{k=1}^{n_p} b_{ik} \, N^p_{k} \gg \alpha_{i}$. 

\begin{figure*}
\begin{center}
\includegraphics[scale=0.4]{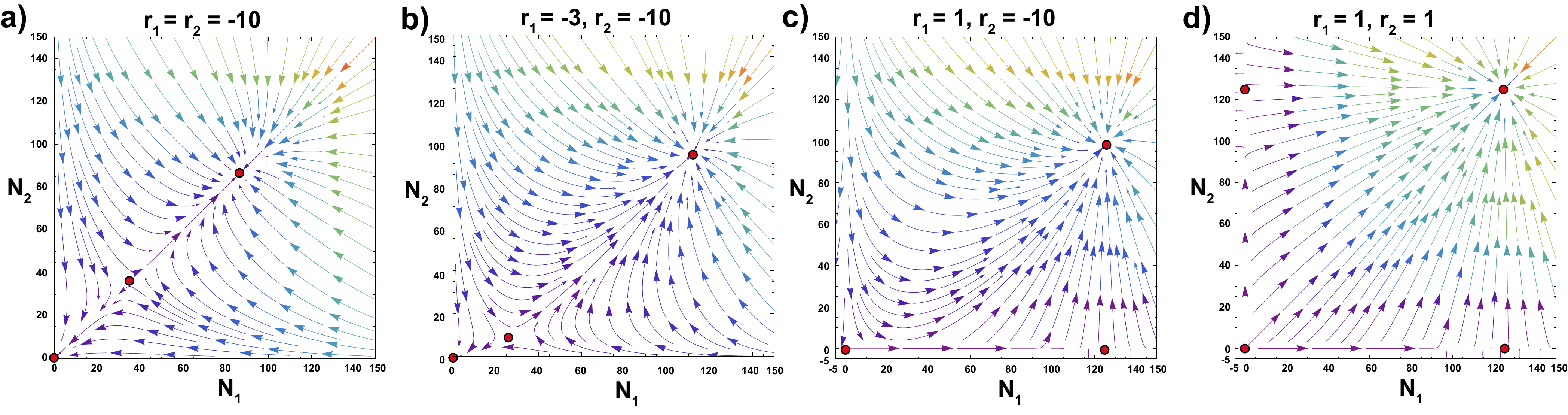}
\caption {Flow diagram for the dynamics of a two species community following the population equations \eqref{eq:dos_especies}. The fixed points are marked as red circles, while the color of the arrows indicates the intensity of the flow. The four panels correspond to different configurations of the intrinsic growth rates $r_1$ and $r_2$. The other parameters of the equations are set at $\alpha_1 = \alpha_2 = 0.008$, $b_{12} = b_{21} = 0.4$ and $c_1 = c_2 = 0.008$. Mutualism is mandatory for both species in a) and b), although in different degree in the diagram b). It is mandatory for species 2 in c), while species 1 can survive without species 2. And, finally, mutualism is facultative for both species in d).}
\label{diagram}
\end{center}
\end{figure*}

\section{Stability Analysis}
\label{stability}  

\subsection{A two species community}

For simplicity, we start the stability analysis by considering a 2-species model for which we can obtain full analytical results. Let the plant species correspond to the index $1$ and the animal species to the index $2$. Equations \eqref{eq:modeloralphaconmut} become then
\begin{align}
\frac{dN^p_{1}}{dt} = \left( r_{1}+ b_{12}\, N^a_{2}\right) \ N^p_{1} - \left(\alpha_{1}+ c_{1} \, b_{12} \, N^a_{2} \right) {N^p_{1}}^2 ,\nonumber\\ 
\frac{dN^a_{2}}{dt} = \left( r_{2}+ b_{21}\, N^p_{1}\right)N^a_{2} - \left(\alpha_{2}+ c_{2} \, b_{21}\, N^p_{1} \right) {N^a_{2}}^2 .
\stepcounter{equation}\tag{\theequation}\label{eq:dos_especies}
\end{align}
Some examples with the flux diagrams for this equation system under different parameter conditions are depicted in Figure \ref{diagram}.

Setting $\frac{dN^p_{1}}{dt} = \frac{dN^a_{2}}{dt} = 0$, one can find the fixed points for the system dynamics. The first, obvious one is total extinction at $({N^p_{1}}^*,{N^a_{2}}^*) = (0,0)$, which is always a fixed point regardless of the parameter values. If any of the intrinsic growth rates $r_1$, $r_2$ is positive, there exist additional fixed points accounting for partial extinctions. The dynamics of the surviving population with positive $r$ follows a decoupled logistic equation, as can be seen from \eqref{eq:dos_especies}. Therefore, its population will tend to the limit given by a non interacting system: Either $K_1 = r_1/\alpha_1$ or $K_2 = r_2/\alpha_2$. This means that there are partial extinction fixed points at $(K_1,0)$ or $(0,K_2)$, or both if mutualism is facultative only for species $1$ ($r_1 >0$), only for species $2$ ($r_2 >0$) (see Figure \ref{diagram}c) or for both ($r_1>0$ and $r_2 >0$) (see Figure \ref{diagram}d), respectively. 

Besides total or partial extinction, other non-trivial fixed points may appear whenever the condition $r_{ef,i} = r_{ef,j} = 0$ is satisfied. At those points, the following relations are fulfilled
\begin{align}
{N^p_{1}}^* = \frac{ r_{1}+ b_{12} \, {N^a_{2}}^* }{\alpha_{1}+ c_{1}\, b_{12}\, {N^a_{2}}^* } , \nonumber\\ 
{N^a_{2}}^* = \frac{ r_{2}+ b_{21}\, {N^p_{1}}^* }{\alpha_{2}+ c_{2} \, b_{21}\, {N^p_{1}}^* } .
\label{eq:puntosfijos}
\end{align}
Substituting the expression for ${N^{a}_2}^*$ on the upper equation, one finds that ${N^p_1}^*$ must satisfy a quadratic equation at the fixed points: 
\begin{equation}
A\, {{N^p_1}^*}^2 + B \, {N^p_1}^* + C=0 ,
\label{eq:quadra}
\end{equation}
where the coefficients $A$, $B$ and $C$ are given by
\begin{align}
\displaystyle A &= c_{2}\, b_{21}\, \alpha_{1}+c_{1}\, b_{12}\, b_{21} , \nonumber \\
\displaystyle B &= \alpha_{1}\, \alpha_{2}+ c_{1}\, b_{12}\, r_{2} - c_{2}\, b_{21}\, r_{1} - b_{12}\, b_{21} ,\nonumber\\
\displaystyle C &= - r _{1}\, \alpha_{2} - b_{12}\, r_{2} .
\label{eq:puntos_n1}
\end{align}
The fixed points of ${N^a_2}^*$ are found by substituting in turn ${N^p_1}^*$ into the bottom expression of Equation \eqref{eq:puntosfijos}. There are several possible scenarios depending on the solutions of Equation \eqref{eq:quadra}:
\begin{enumerate}
\item Both roots are complex. There are no additional fixed points, except for total or partial extinction.
\item A unique real root. This is a bifurcation point for the system dynamics, the solutions are real but degenerate. In this case, there exists a single fixed point besides extinction. The final system fate depends on the stability of this point. However, the most likely outcome is that the populations get eventually extinct. 
\item Both roots are real and different. The situation is similar to the one displayed in Figure~\ref{diagram}a. There are two non-trivial fixed points, typically one stable, and one saddle points that lies on the boundary between two attraction basins. The position of the saddle point determines the extension of the extinction basin and, therefore, the resilience of the system to external perturbations. We call this point the \emph{extinction threshold} and its position will be denoted by $({N_{1}^{p}}^\bullet,{N_{2}^{a}}^\bullet)$.
\end{enumerate}

In order to study the linear stability of the fixed points, we can expand the Equations \eqref{eq:dos_especies} in a Taylor series around them and calculate the Jacobian of the system (see  Appendix A for details). If the eigenvalues are negative, the fixed point is stable. Otherwise, it can be a saddle point if one is positive and the other negative or unstable if both are negative. Starting by total extinction, the Jacobian can be written as 
\begin{equation}
J = \left(
\begin{array}{ll}
r_{1}   & 0 \\
0 & r_{2} 
\end{array}
\right) .\stepcounter{equation}\tag{\theequation}\label{eq:J00}
\end{equation}
The eigenvalues are $\lambda_{1,2} = r_{1,2}$, which means that the extinction point is linearly stable under the assumption of $ r_{1}<0$ and $r_{2}<0$, i.e. when both species rely on mutualism for survival. Total extinction has in this case an attraction basin for different population values. If the system falls within this population levels, the only possible fate is extinction. 

On the other hand, if mutualism is facultative for one or both species, total extinction becomes a saddle or unstable point. However, other two fixed points can appear for partial extinction. In this case, the condition for stability of $(r_1/\alpha_1, 0)$ is that $r_{1}>0$ and $r_{2}<-b_{21}\, r_{1}/\alpha_{1}$. Similarly, $(r_1/\alpha_1, 0)$ is stable only if  $r_{2}>0$ and $r_{1}<-b_{12}\, r_{2}/\alpha_{2}$.

The same analysis for the remaining non-trivial fixed points leads us to the Jacobian matrix: 
\begin{equation}
J = \left(
\begin{array}{ll}
- {N^{p}_{1}}^* \, (\alpha_{1}+ c_{1}\, b_{12} \, {N^a_2}^* )  & {N_{1}^{p}}^* \, b_{12} \, (1 - c_{1}\, {N_{1}^{p}}^* ) \\
{N_{2}^{a}}^* \, b_{21}\, (1 - c_{2}\, {N_{2}^{a}}^* ) & - {N_{2}^{a}}^* \, (\alpha_{2}+ c_{2}\,b_{21}\,{N_{1}^{p}}^* )
\end{array}
\right)\stepcounter{equation}\tag{\theequation}\label{eq:J}
\end{equation}
Since the parameters $c_{1}$ and $c_2$ are always positive (remember that they are the inverse of the maximum population in the limit of strong mutualism), all the terms of $J$ have the sign shown in Equation \eqref{eq:J}. The diagonal terms are negative, while the off-diagonal are always positive. A similar configuration for the Jacobian matrix  was observed in mutualistic models in \cite{goh79}. It implies that the eigenvalues of $J$ are both real and that they can be either both negative (\emph{stable fixed points}) or one positive and another negative (\emph{saddle point}). The condition for the existence of a \emph{saddle point} is that the determinant of the Jacobian matrix at the \emph{extinction threshold} is negative, $J_{11} \, J_{22} < J_{12}\, J_{21}$, which in terms of ${N_{1}^{p}}^\bullet$ and  ${N_{2}^{a}}^\bullet$ means that
\begin{equation}
1-c_{1}\, {N_{1}^{p}}^\bullet - c_{2}\, {N_{2}^{a}}^\bullet > 0 .
\end{equation}

\begin{figure}
\centering
\includegraphics[scale=0.6]{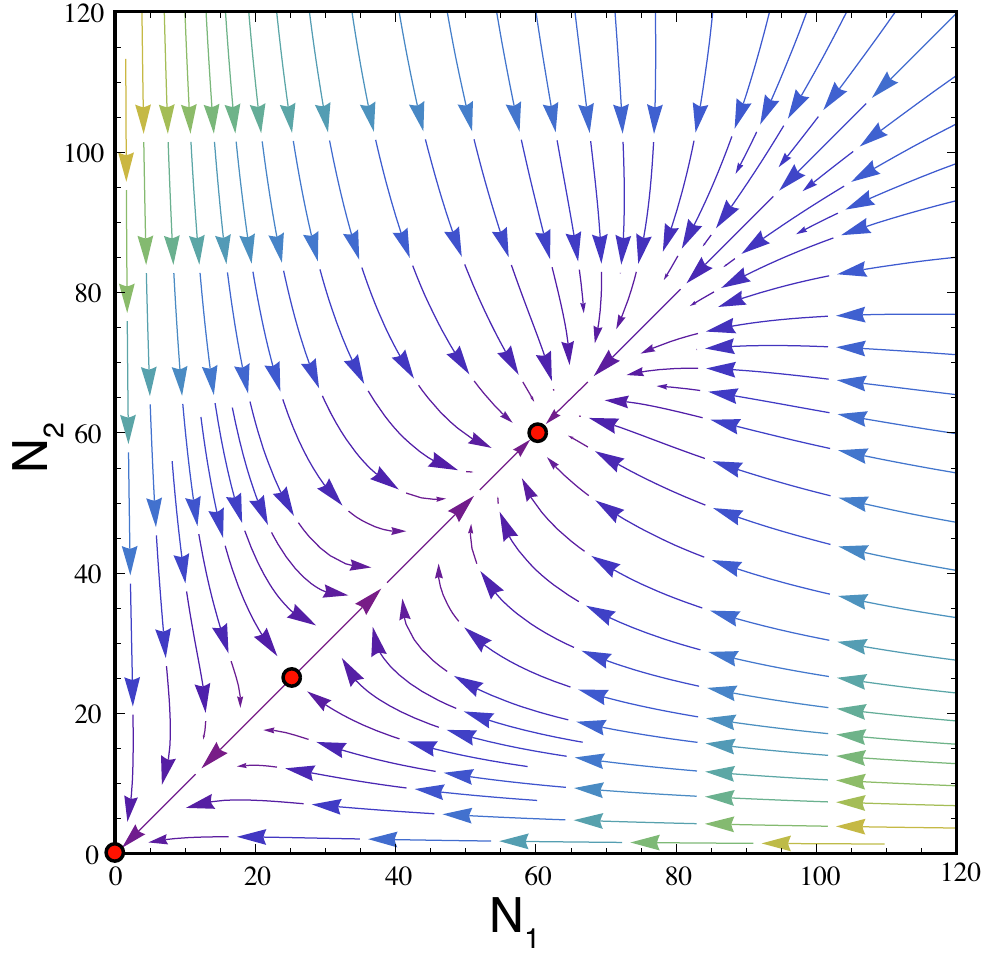}
\caption {Flow diagram for the dynamics of the type II Equations \eqref{eq_typeII}. The fixed points are marked as red circles, while the color of the arrows indicate the intensity of the flow. Finding this dynamical configuration took a considerable effort in parameter tuning. The equation parameters used here are $r_1 = r_2 = -0.1$, $\alpha_1 = \alpha_2 = 0.001$, $a = 0.066$, $b = 0.2$ and $T_H = 1$.}
\label{typeII}
\end{figure}

All these results for two species show that our model displays a rich dynamics. Still, it is simple enough to understand well its different regimes and where they appear in the parameter space. In this sense, it overcomes shortcomings inherent to the type II formulation. For instance, finding a dynamic configuration as the one shown in Figure~\ref{typeII} for the model of Equation \eqref{eq_typeII} requires a notable effort in terms of parameter tuning. This dynamical configuration with two attractors and a saddle point is ideal to study issues such as system resilience, capacity to bear a high biodiversity or the evolution of the mutualistic interaction networks (see, for example, \cite{bastolla09} or \cite{suweis13}). Such regime appears naturally in our model, as in Figure~\ref{diagram}a, without the need of an elaborated parameter search.

\subsection{Survival watershed}
\label{watershed}

We will refer as \emph{survival watershed} to the repeller limit between trajectories that evolve towards full system capacity or towards extinction. In Figure \ref{diagram}a, it corresponds to the curve delimiting the attraction basin of total extinction. The watershed includes the non-trivial saddle point $({N_1^p}^\bullet,{N_2^a}^\bullet)$. Its location in the phase space is important because it determines the fragility or robustness of the system by establishing the extension of the extinction basin. Some characteristics of the points laying on the watershed can be analytically found at least for the case of two species communities. The points of the watershed correspond to population pairs $({N_1^p},{N_2^a})$ for which the system dynamics remains in the watershed and ends at $({N_1^p}^\bullet,{N_2^a}^\bullet)$. 

By definition, at $({N_1^p}^\bullet,{N_2^a}^\bullet)$ both effective growth rates are zero. To reach this point from any other initial populations, the effective growth rates of both species need to have different signs and evolve similarly in time. If both had the same sign (positive or negative), the system dynamics would be attracted towards full capacity or towards total extinction. Let us assume that the system is approaching $({N_1^p}^\bullet,{N_2^a}^\bullet)$, that the initial populations were $({N_1^p}^0,{N_2^a}^0)$ on the watershed and that we can write the effective growth rates as
\begin{align}
r_{ef,1}  = & \, A \, e^{-\gamma\, t} ,\nonumber\\  
r_{ef,2}  = & -B\, e^{-\gamma\, t} , 
\label{eq:coeffsreffs}
\end{align}
where $A$, $B$ and $\gamma$ are constants, unknown at the moment. Equations \eqref{eq:dos_especies} then become
\begin{align}
\frac{dN^p_{1}}{dt} & = N^p_{1} \, A \, e^{-\gamma \, t} , \nonumber \\
\frac{dN^a_{2}}{dt} & = -N^a_{2}\, B \, e^{-\gamma \, t} .
\label{eq:coeffsreffs_2}
\end{align}
Integrating these equations between $t = 0$ and infinity, we find that
\begin{align}
 \ln \frac{{N_1^p}^\bullet}{{N_{1}^p}^0} & = \frac{A}{\gamma} , \nonumber\\ 
 \ln \frac{{N_2^a}^\bullet}{{N_{2}^a}^0} & = - \frac{B}{\gamma} .
\label{eq:coeffsreffs_3}
\end{align}
Equating the value of $\gamma$ in both expressions, we get the condition for $({N_1^p}^0,{N_2^a}^0)$ to be part of the survival watershed:
\begin{align}
\frac{1}{B} \ln \left(\frac{{N_{2}^a}^\bullet}{{N_{2}^a}^0} \right) + \frac{1}{A} \ln \left(\frac{{N_1^p}^\bullet}{{N_1^p}^0} \right) = 0 ,
\label{eq:coeffsreffs_4}
\end{align}
which means that the functional form of the watershed is given by the power-law 
\begin{align}
{N_2^a}^0 = C\, ({N_1^p}^0)^\frac{-B}{A}. 
\label{eq:powerlaw}
\end{align}
$C$ is a constant that taking into account that the watershed includes the fixed point $({N_1^p}^\bullet,{N_2^a}^\bullet)$ can be written as
\begin{align}
C = {N_2^a}^\bullet / ({N_1^p}^\bullet)^\frac{-B}{A} .
\end{align}

\begin{figure}
\centering
\includegraphics[scale=0.27]{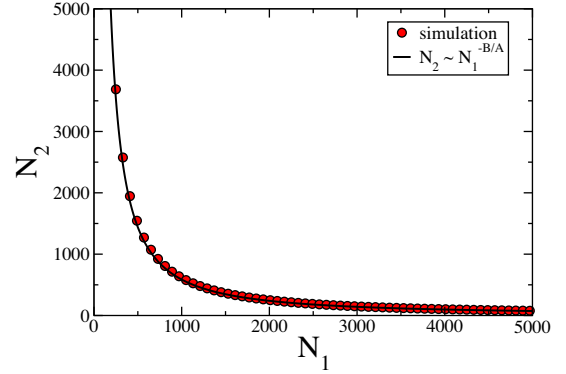}
\caption {Survival watershed for two species. Dots were found performing a numerical scan of the system dynamics, determining for which initial conditions the final outcome was extinction or full capacity. Grey solid line is the power law found with Equations \eqref{eq:powerlaw} and \eqref{eq:implicita_watershed}. In this case, $\frac{B}{A}=1.2944,~{N_1^p}^\bullet=989,~{N_2^a}^\bullet=1232,~b_{12}=0.000041850,~c_{1}=0.00004,~\alpha_{1}=~0.000035,~r_1=-0.016,~b_{21}=0.00008750,~c_{2}=0.0001,~\alpha_{2}=0.000035,~r_2 =-0.02$.}
\label{fig:powerlaw}
\end{figure}

To find the value of the exponent $\frac{B}{A}$, we must return to the definition of the effective growth rates, $r_{ef,1}$ and $r_{ef,2}$. According to Equations \eqref{eq:coeffsreffs}, at $t=0$ we have that 
\begin{align}
A = & \, r_{1}+ b_{12}\, {N_2^a}^0 - (\alpha_{1}+ c_{1} \, b_{12}\, {N_{2}^a}^0) \, {N_1^p}^0 , \nonumber\\
-B = &\, r_{2} + b_{21} \, {N_{1}^p}^0-(\alpha_{2}+ c_{2}\,  b_{21}\, {N_{1}^p}^0)\,  {N_{2}^a}^0 .
\label{eq:reffs_2especies}
\end{align}
If we know that our initial points are part of the watershed, dividing these two expressions we can obtain the exponent value. Alternatively, if other points in the watershed apart from $({N_1^p}^\bullet,{N_2^a}^\bullet)$ need to be found, we can divide the previous expressions, one by the other, and using Equation \eqref{eq:coeffsreffs_3} reach the following implicit equation
\begin{align}
\frac {\ln \left( \frac{{N_2^a}^\bullet}{{N_2^a}^0} \right)}{\ln \left( \frac{{N_1^p}^\bullet}{{N_1^p}^0} \right)} = \frac{( r_{2}+ b_{21}\, {N_1^p}^0) - (\alpha_{2}+ c_{2} \,  b_{21}\, {N_1^p}^0 ) \, {N_1^p}^0}{( r_{1}+ b_{12}\, {N_2^a}^0) - (\alpha_{1}+ c_{1} \, b_{12}\, {N_2^a}^0 ) \, {N_2^a}^0 } .
\label{eq:implicita_watershed}
\end{align}
Solving then numerically this equation we can find other points in the watershed and with them an estimation of $\frac{B}{A}$. Figure \ref{fig:powerlaw} shows an example of the watershed and a comparison between the curve obtained with Equations \eqref{eq:powerlaw} and \eqref{eq:implicita_watershed} and numerical estimations integrating the system dynamics.

\subsection{General communities}
 
The generalization of the stability analysis for an arbitrary number of species is straightforward. The fixed points of Equations~\eqref{eq:modeloralphaconmut} comprise the trivial solution $(N_{i}^p,\cdots, N_{j}^a) = (0, \cdots,0)$, i.e., total extinction, \emph{partial extinction} points if mutualism is facultative for any species, and the populations $({N^{a}_{i}}^*,\cdots,{N^{p}_{j}}^*)$ for which the effective growth rates vanish:
\begin{align}
r^{*}_{ef,i}  = (r_{i}+ \sum_{k=1}^{n_{p}}\,  b_{ik}\, {N^p_{k}}^*)- (\alpha_{i}+c_{i}\, \sum_{k=1}^{n_{p}} b_{ik}\, {N^{p}_k}^* )\, {N^{a}_{i}}^* = 0 \nonumber ,\\
r^{*}_{ef,j}  = (r_{j}+ \sum_{\ell=1}^{n_{a}} b_{j\ell}\, {N^{a}_{\ell}}^*)- (\alpha_{j}+c_{j}\, \sum_{\ell=1}^{n_{a}} b_{j\ell}\, {N^{a}_\ell}^* )\, {N^{p}_{j}}^* 
=0 ,
\label{eq:effrate2}
\end{align}
for animals and plants, respectively. These equations can be rewritten as
\begin{align}  
{N^{a}_{i}}^* = \frac{r_{i}+\sum_{k=1}^{n_{p}}b_{ik}\, {N^{p}_{k}}^*}{\alpha_{i}+c_{i}\,\sum_{k=1}^{n_{p}}{b_{ik}N^{p}_{k}}^*} = 
  \frac{r_{i}+r_{i}^{mut}}{\alpha_{i}+c_{i}\, r_{i}^{mut}} = 
  \frac{r_{i}^{*+}}{r_{i}^{*-}} , \nonumber\\
{N^{p}_{j}}^*=\frac{r_{j}+\sum_{\ell=1}^{n_{a}}b_{j\ell}\, {N^{a}_{\ell}}^*}{\alpha_{j}+c_{j}\,\sum_{\ell=1}^{n_{a}}{b_{j\ell}N^{a}_{\ell}}^*} =
  \frac{r_{j}+r_{j}^{mut}}{\alpha_{j}+c_{j}r_{j}^{mut}} =
  \frac{r_{j}^{*+}}{r_{j}^{*-}} .
\end{align} 
The rates $r_{i}^{mut}$ account for the effect of the mutualism on species $i$, while the rates $r^{*+}$ stand for the terms increasing the population growth and $r^{*-}$ for those decreasing it via intra-specific competition. 

Equations \eqref{eq:modeloralphaconmut} can be linearized around the fixed points. The corresponding Jacobian matrix has the same appearance as its counterpart for a two species community (Equation \eqref{eq:J}), with negative entries on the diagonal and positive (and null) entries for the off-diagonal elements. For the non-trivial fixed points (those without total or partial extinctions), the diagonal terms can be written for animals and plants, respectively, as (see Appendix A)
\begin{align}
\displaystyle & J_{ii}= - {N^{a}_{i}}^* \left(\alpha_{i} + c_{i} \,  \sum_{k=1}^{n_{p}} b_{ik} {N^{p}_{k}}^* \right), \nonumber\\
\displaystyle & J_{jj}= - {N^{p}_{j}}^* \left(\alpha_{j} + c_{j} \, \sum_{\ell=1}^{n_{a}} b_{j\ell}\, {N^{a}_{\ell}}^*\right).
\label{eq:Jii}
\end{align}
The non-diagonal terms, in turn, are
\begin{align}
\displaystyle & J_{ij}={N^{a}_{i}}^* \, b_{ij}\, \left( 1-c_{i}\, {N^{a}_{i}}^*\right) 
\label{eq:Jij1}
\end{align}
for interactions between a generic animal species $i$ and a plant $j$, and 
\begin{align}
\displaystyle & J_{ji}={N^{p}_{j}}^* \, b_{ji}\, \left( 1-c_{j}\, {N^{p}_{j}}^*\right)
\label{eq:Jij2}
\end{align}
for the opposite  interactions between plant $j$ and animal $i$. Given the invariance of the trace of a matrix to change in the vector basis, the sum of the eigenvalues of the Jacobian matrix must satisfy the relation
\begin{equation}
  \sum_{k}^{n_{a}+n_{p}} \lambda_{k}= - \left(\sum_{k}^{n_{a}+n_{p}} |J_{kk}| \right) .
  \stepcounter{equation}\tag{\theequation}\label{eq:sum_lambdas2}
\end{equation}
The trace is negative, which means that if there are any positive or null eigenvalues their effect must be compensated by several other negative eigenvalues. Therefore, the non-trivial fixed points can be either stable (if all the eigenvalues are negative) or saddle points, if at least one is positive. They cannot be purely unstable.

Another question to discuss is what occurs in case of partial extinctions. The effect of the extinction of some species in the system is to reduce the dimensionality of the set of Equations \eqref{eq:modeloralphaconmut}. To fix ideas, let us assume, for instance, that one animal species $e$ gets extinct. This implies that the possible fixed points for the system dynamics must include now that ${N_e^a}^* = 0$. The collapse of $e$ can trigger the extinction of some plant species relying on it for reproduction. After these plants, some other animals depending on them can in turn get extinct, and so on forming a cascade extinction event. Note that, although the extinction event can be produced by external factors to the system such as a disease or a famine, the population dynamics for the remaining species is linked to the full system equations. The new non-trivial fixed points correspond to the partial extinction points of the original complete set of equations. The stability of these points can substantially change. The entries of the Jacobian matrix for the extinct species in the new non-trivial fixed points become $J_{ee} = r_e + \sum_{k =1}^{n_p }b_{ek}\,{N_k^p}^*$ in the diagonal and  $J_{ej} = 0$ off the diagonal. These terms do not contribute to eigenvalues relevant for the stability analysis. The rest of entries for the Jacobian are given by Equations \eqref{eq:Jii}, \eqref{eq:Jij1}, and \eqref{eq:Jij2} adapted to the surviving species. This means that the sums of Equations \eqref{eq:Jii} do not run over all the species as before, and that the diagonal terms can be closer to zero. The stability of the new fixed points can thus change depending on the parameters of the equations ruling the population dynamics of the surviving species. Actually, depending on how the interactions between species are in the remaining community, the system can become more robust to external perturbations after a partial extinction event.   

\section{Numerical results}
\label{results}

The previous analytical results are general so can be used in any mutualistic community. However, to fix ideas, it is important to focus on a particular example. To be able to follow the system dynamics, a numerical technique to integrate the Equations~\ref{eq:modeloralphaconmut} is implemented. We have used a stochastic approach to take into account the discrete nature of the individuals in a population. A similar technique has been applied before to epidemiologic studies (see, for instance, \cite{balcan2009multiscale}). Details on the model implementation are given in \ref{NumSim}.

\begin{figure}
\centering
\includegraphics[scale=1.1]{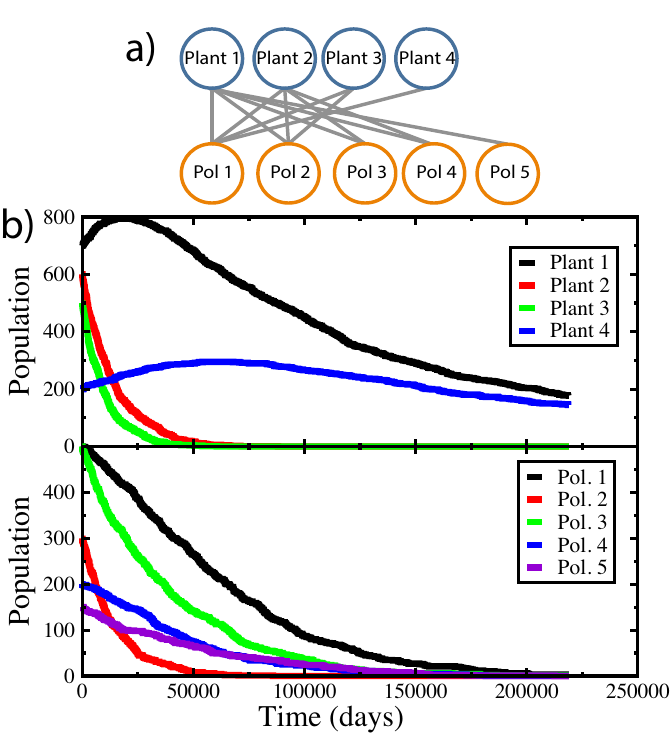}
\caption {a) Mutualistic community with four species of plants and five species of pollinators. b) Simulation results with the population trends for the different species (each species is color-coded). Numerical solution shows that initial populations are below the \emph{extinction threshold}. In this scenario, the system tends to total extinction. The parameters of the simulation can be found in the \ref{DataTables} and table \ref{tab:experiment1}.}
\label{fig:red_exper_stab1}
\end{figure}

The intrinsic growth rates are fixed in negative values for all the simulations, which implies that mutualism is always obligatory for all species. Figure \ref{fig:red_exper_stab1}a shows a small mutualistic community created for the purpose of this analysis (see numerical details in \ref{DataTables}, the simulations parameters are in Table \ref{tab:experiment1}). In many empirical studies, the number of interacting species in each class is of the order of tens. The network of this example has less species but already displays the main behaviors of larger communities. The population dynamics for the first simulation is depicted in Figure \ref{fig:red_exper_stab1}b. The conditions are such that seven out of nine species have negative effective growth rates. This leads to a decrease in all the populations except in those of plant species $1$ and $4$. Still, despite their initial growth, the decline of their mutualistic partners turn negative their $r_{ef,i}$ and they eventually get extinct. This scenario shows how the system is attracted to extinction if the populations that are initially below the extinction threshold.

\begin{figure}
\centering
\includegraphics[scale=0.35]{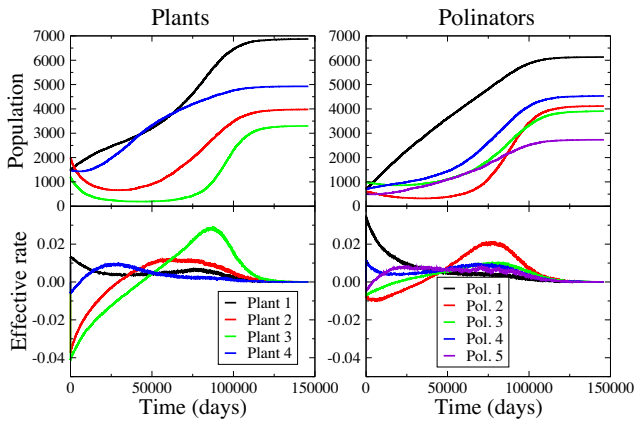}
\caption {Population dynamics and evolution of effective rates for the different species (each species is color-coded). The interaction network is the same as in Figure~\ref{fig:red_exper_stab1}a. Despite the initial negative effective growth rates for some species, the system dynamics in this scenario tends to full capacity. The numerical details on the simulation are in \ref{DataTables}, table \ref{tab:experiment2}.}
\label{fig:exper_stab2}
\end{figure}

The next simulation explores other fixed points of the model dynamics. Again, all intrinsic rates are negative but mutualistic interaction weights (terms $b_{ij}$) and initial populations are selected in such a way that the effective growth rates of plants $1$ and pollinators $1$ and $4$ are positive, while the effective growth rates of all the other species are negative (see Table B.2). Despite this initial disadvantage, the population of these species recover and the system dynamics tends to the fixed point at full capacity (Figure ~\ref{fig:exper_stab2}). The speed of the recovery process is different for all the species and even for some of them there is an initial decline of the population size. This short time tendencies can deceive an observer unless the observation period is long enough to comprehend the full system dynamics.

\begin{figure}
\centering
\includegraphics[scale=0.9]{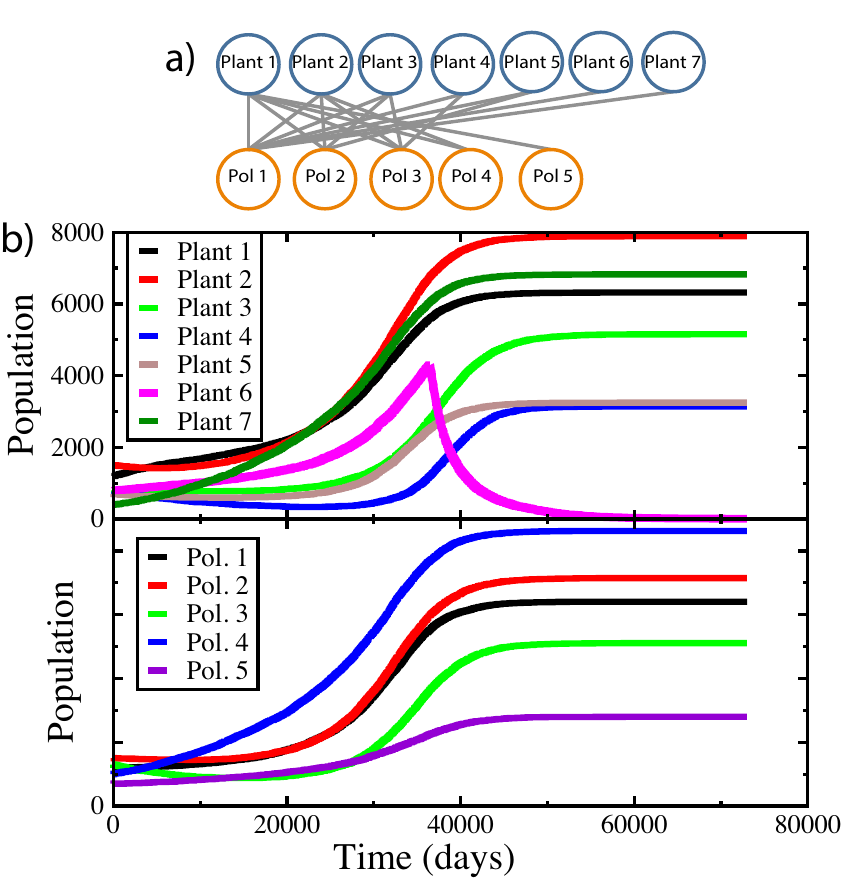}
\caption { a) A plant-pollinator network with high nestedness. b) Simulation results of population trends obtained with this network. An external perturbation attacks plant species 7, which leads to its extinction. The rest of the community reaches a stationary state at full capacity in the reduced system. Numerical details on the simulation are in \ref{DataTables}, including the simulation parameters in Table \ref{tab:exper_resilience_strong}.}
\label{fig:exper_resilience_strong}
\end{figure}

System stability analysis are usually performed under the assumption of  constant external conditions. However, these conditions may strongly vary in more realistic scenarios due to factors such as diseases, famines or droughts. The resilience of mutualistic networks and foodwebs has been traditionally related to a network property named \emph{nestedness} \citep{bascompte2003nested}. Two types of species can be found in interaction networks:  \emph{generalists}, linked to several instances of the other class, and \emph{specialists}, tied only to a small number of them. In nested networks, there is a core of generalist species that are highly coupled, whereas specialists are much more likely to be connected to generalists than to other specialists. Specialists can suffer more in an adverse scenario, but the core of generalist is able to sustain the community. In the next numerical simulations, we explore the effect of nestedness on the system resilience using our model. The objective is to explore whether its dynamics responds similarly to an increase in the nestedness level of the interaction network. These are simple examples but they already help to fix ideas.

In the first example, a network with seven species of plants and five of pollinators is considered (Fig. \ref{fig:exper_resilience_strong}a). We are not going to develop a formal justification, but this network is strongly nested with an easy to identify core of \emph{generalist} species and \emph{specialists} tied to \emph{generalists} of the other class. Initial populations have been chosen to be above the survival threshold. The system is evolved until it reaches population capacity until year $100$ (day $36500$, see Fig. \ref{fig:exper_resilience_strong}b). Then, a disruption is introduced in the form of plague attacking  plant species 6. This plant suffers an additional $0.20$  yearly death rate and it becomes extinct. Plant species $6$ is only linked to pollinator species $1$, the most generalist of its class. The effect of its extinction is negligible since mutualistic benefit of the rest of plant species is high enough to balance it.

\begin{figure}
\centering
\includegraphics[scale=0.9]{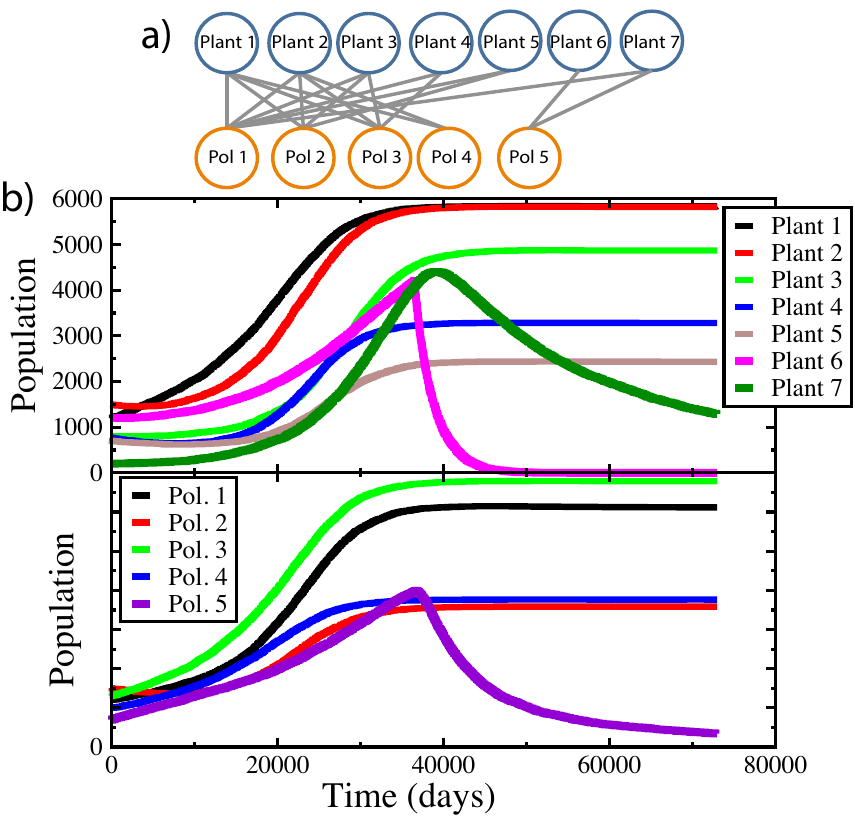}
\caption { a) A low nested interaction network. b) Simulation results depicting population trends in the low nested network. As before, an external perturbation attacks plant species 6. The system, however, does not recover and a small scale extinction event is triggered. Numerical details of the simulation can be found in \ref{DataTables}, the simulation parameters are included in Table \ref{tab:exper_resilience_weak}.}
\label{fig:red_exper_resilience_weak}
\end{figure}

In the last example, a slight modification of the network is used (Fig. \ref{fig:red_exper_resilience_weak}a), that breaks the strong nestedness of previous example. This time plant species $6$ is linked to pollinator species $5$, an specialist. We also remove the link connecting plant $1$ and pollinator $5$ and add a link between plant $7$ and pollinator $5$. Numerical values of the rates for the interaction network are described in \ref{DataTables}, Table \ref{tab:exper_resilience_weak}. The simulation is then repeated but this time with less nested network. All initial effective rates eventually turn positive by the growth of the system. At year 100, plant species $6$ suffers the same attack as before, an additional $0.20$ yearly death rate, that triggers its extinction. However, the effect this time is  different. Pollinator species $5$ depends for its survival on plant species $6$, so the slope of its population becomes negative and will eventually vanish. Plant $7$, connected to specialist pollinator $5$ and with a weak tie with pollinator $1$, losses its main source of mutualistic benefit and also faces extinction. So, an external event on plant $6$ has dragged plant $7$ to extinction because they were indirectly linked by specialist pollinator $5$. If both plant species share links with a generalist pollinator this cascade effect is more unlikely.

\section{Conclusions}
\label{discuss}

In this work, we have introduced a model derived from the logistic approach to study population dynamics under mutualistic interactions. The proposed equations overcome the drawbacks of May's model when dealing with negative growth rates, an important issue when the system is far from equilibrium and mutualism is obligatory. Our model also allows for an easier analytical treatment since the nonlinearities are simpler than for instance those of the type II models. This simplicity makes it also easier to estimate from empirical data the different rates involved in the equations or to assign them an ecological interpretation.  This is a key point because empirical mutualistic interaction datasets are scarce since its compilation is a painstaking task.

We have studied the dynamics of the model finding the dynamics fixed points and their stability analytically for a simple case, and numerically for a more involved community. Our model shows the fixed point structure of May's model with the notable addition of a saddle point that controls the stability of the whole system. In this regard the model is as rich in dynamic behaviors as the type II models but with a much simpler mathematical structure. We have analyzed numerically the resilience of our model to external perturbations introducing perturbations in a simple but relatively involved mutualistic network. As in other communities described in the literature, the system resilience is a function of the structure of the network. We hope that this new model can be used to gain further insights in the mutualistic communities due to its rich dynamics and simplicity.



\section*{Acknowledgements}

We have received partial financial support from the Spanish Ministry of Economy (MINECO) under projects MTM2012-39101, MODASS (FIS2011-24785), LIMITES (CGL2009-07229), and AdAptA(CGL2012-33528); from the project PGUI of Comunidad de Madrid MODELICO-CM/S2009ESP-1691 and from the EU Commmission through projects EUNOIA and LASAGNE. JJR acknowledges funding from the Ram\'on y Cajal program of MINECO.

\appendix

\section{Detailed Linear Stability Analysis}

For sake of simplicity, we drop the use of the superscripts for plants and animals. The equations system (\ref{eq:dos_especies}) can be expanded in a Taylor series around the singular point ($N^{*}_{1}, N^{*}_{2}$) as  $N_{1}= N^{*}_1+\tilde{N}_{1}$ and $N_{2}= N^{*}_2+\tilde{N}_{2}$ \citep{murraymathematical}:
\small
\begin{equation}
\begin{array}{lcr}
\displaystyle \frac{d\tilde{N}_{1}}{dt} = r_{1}+ b_{12}(N^{*}_2+\tilde{N}_{2})-(\alpha_{1}+ c_{1} b_{12} (N^{*}_2+\tilde{N}_{2}))( N^{*}_1+\tilde{N}_{1})\nonumber\\
\\
\displaystyle \frac{d\tilde{N}_{2}}{dt} = r_{2}+ b_{21}( N^{*}_1+\tilde{N}_{1})-(\alpha_{2}+ c_{2} b_{21}(N^{*}_1+\tilde{N}_{1}))(N^{*}_2+\tilde{N}_{2}) 
\stepcounter{equation}\tag{\theequation}\label{eq:effrateTaylor}
\end{array}
\end{equation}
\normalsize

\noindent and retaining only the linear terms we get:
\small
\begin{equation}
\begin{array}{lcr}
\displaystyle \frac{d\tilde{N}_{1}}{dt}= \tilde{N}_{2}( b_{12} - c_{1} b_{12}\, N^{*}_1)-\tilde{N}_{1}(\alpha_{1}+ c_{1} b_{12} \, N^{*}_2) \equiv f_{1}(\tilde{N}_{1},\tilde{N}_{2}) \nonumber\\
\\
\displaystyle \frac{d\tilde{N}_{2}}{dt} = \tilde{N}_{1}( b_{21} - c_{2} b_{21}\, N^{*}_2)-\tilde{N}_{2}(\alpha_{2}+ c_{2} b_{21} \, N^{*}_1) \equiv f_{2}(\tilde{N}_{1},\tilde{N}_{2})
\stepcounter{equation}\tag{\theequation}\label{eq:effrateTaylor2}
\end{array}
\end{equation}
\normalsize

The Jacobian matrix entries are:

\begin{equation}
\begin{array}{l}
J_{11}= \frac{\partial f_{1}}{\partial \tilde{N}_{1}} =- N^{*}_{1}\left(\alpha_{1}+ c_{1} b_{12} \, N^{*}_2\right)  \\
\\
J_{12}= \frac{\partial f_{1}}{\partial \tilde{N}_{2}} = N^{*}_{1}b_{12}\left(1 - c_{1}\, N^{*}_1\right) \\
\\
J_{21}= \frac{\partial f_{2}}{\partial \tilde{N}_{1}} = N^{*}_{2}b_{21} \left(1 - c_{2}\, N^{*}_2\right) \\
\\
J_{22}= \frac{\partial f_{2}}{\partial \tilde{N}_{2}} = - N^{*}_{2}\left(\alpha_{2}+ c_{2} b_{21}\,N^{*}_{1}\right)
\end{array}
\stepcounter{equation}\tag{\theequation}\label{eq:J11}
\end{equation}

\noindent and it can be written in terms of positive entries $J_{ij}$ as

\begin{equation*}
J = \left(
\begin{array}{rr}
-J_{11} & J_{12} \\ J_{21} & -J_{22}
\end{array}
\right)
\end{equation*}

The eigenvalues $\lambda_{1,2}$ can be obtained from:
\begin{equation}
\lvert J - \lambda I \rvert =0
\stepcounter{equation}\tag{\theequation}\label{eq:lambda0App}
\end{equation}

\noindent whose solutions are
\begin{equation}
\begin{array}{lcl}
\lambda _{1,2}=\frac{1}{2}\left(tr(J)\pm \sqrt{tr^{2}(J)-4\,\mathrm{Det}(J)}\right)\\ =
\frac{1}{2}\left(-\left(J_{11}+J_{22}\right)\pm \sqrt{\left(J_{11}+J_{22}\right)^{2}-4\,\mathrm{Det}(J)}\right)\\  =
\frac{1}{2}\left(-\left(J_{11}+J_{22}\right)\pm \sqrt{\left(J_{11}-J_{22}\right)^{2} +4 \,\left( J_{12}J_{21} \right) }\right)
\end{array}
\stepcounter{equation}\tag{\theequation}\label{eq:lambda12}
\end{equation}

The last expression indicates that the two eigenvalues are real.
In addition, eigenvalues satisfy:
\begin{equation}
\prod_{k}\lambda_{k}=\mathrm{Det}(J)
\end{equation}

\noindent so the singular point will be a
\emph{saddle point} when  $\mathrm{Det}(J)<0$.
Expanding the determinant of the Jacobian matrix we obtain a condition for the
singular point:

\begin{equation}
1-c_{1}N^{*}_{1}-c_{2}N^{*}_{2} >0
\end{equation}

The partial extinctions are also singular points, and correspond to $N^{*}_{1,2}=0$. For the sake of simplicity, we only write the equations for the singular point ($N^{*}_{1}=r_{1}/\alpha_{1},N^{*}_{2}=0$). With the Taylor expansion around this point the system equations can be written:

\small
\begin{equation}
\begin{array}{ll}
\displaystyle \frac{d\tilde{N}_{1}}{dt} = & r_{1} N^{*}_{1}-\alpha_{1}N^{*2}_{1}+r_{1}\tilde{N}_{1}+ b_{12}\tilde{N}_{2}N^{*}_1-2\alpha_{1}N^{*}_1\tilde{N}_{1} + \\
\, & - c_{1} b_{12}\tilde{N}_{2}N^{*2}_1\nonumber\\
\displaystyle \frac{d\tilde{N}_{2}}{dt} = & r_{2}\tilde{N_{2}}+ b_{21} N^{*}_1\tilde{N_{2}} 
\end{array}
\label{eq:effrateTaylorN2=0}
\end{equation}
\normalsize

The Jacobian is now

\begin{equation*}
J = \left(
\begin{array}{rr}
-r_{1} & b_{12}N^{*}_{1}\left(1-c_{1}N^{*}_{1}\right) \\
0 & r_{2}+b_{21}N^{*}_{1}
\end{array}
\right)
\end{equation*}

The eigenvalues are the diagonal entries.
This singular point will be a stable node when $r_{1}>0$ and $r_{2}<-b_{21}r_{1}/\alpha_{1}$. The symmetric solution is ($N^{*}_{1}=0,N^{*}_{2}=r_{2}/\alpha_{2}$)
and it will be a  stable node when $r_{2}>0$ and $r_{1}<-b_{12}r_{2}/\alpha_{2}$.

The generalization for $n_{a} + n_{p}$ species is
 
 \begin{align}
\frac{dN_{i}}{dt} = \left( r_{i}+ \sum_{j=1}^{n_{a}} b_{ij}N_{j}\right)N_{i} - \left(\alpha_{i}+ c_{i} \sum_{j=1}^{n_{a}} b_{ij}N_{j} \right) N^{2}_{i} \nonumber\\ 
\frac{dN_{j}}{dt} = \left( r_{j}+ \sum_{i=1}^{n_{p}} b_{ji}N_{i}\right)N_{j} - \left(\alpha_{j}+ c_{j} \sum_{i=1}^{n_{p}} b_{ji} N_{i} \right) N^{2}_{j} 
\stepcounter{equation}\tag{\theequation}\label{eq:N_especies}
\end{align}

\noindent where the subscript $i$ runs for all plant species and the subscript $j$
runs for all animal species.
 
The singular points of this set of equations are: the trivial solution ($N_{i=1\cdots n_{p}}=0, N_{j=1\cdots n_{a}}=0$), i.e. the total extinction point, and the solution of \emph{effective growth rates} equal to zero:

\small
\begin{equation}
\begin{array}{lcr}
\displaystyle r^{*}_{ef,i} =\left(r_{i}+ \sum_{j=1}^{n_{a}} b_{ij}N^{*}_{j}\right)- \left(\alpha_{i}+c_{i}\sum_{j=1}^{n_{a}} b_{ij}N^{*}_j\right)N^{*}_{i}
=0 \nonumber\\
\displaystyle r^{*}_{ef,j} = \left(r_{j}+ \sum_{i=1}^{n_{p}} b_{ji}N^{*}_{i}\right)- \left(\alpha_{j}+c_{j}\sum_{i=1}^{n_{p}} b_{ji}N^{*}_i\right)N^{*}_{j} 
=0 
\stepcounter{equation}\tag{\theequation}\label{eq:effrateN}
\end{array}
\end{equation}
\normalsize

\noindent that can be rewritten as an implicit equation set.

\begin{eqnarray}
\begin{array}{lcc}
  N^{*}_{i}=\frac{r_{i}+\sum_{j=1}^{n_{a}}b_{ij}N^{*}_{j}}{\alpha_{i}+c_{i}\sum_{i=1}^{n_{p}}b_{ij}N^{*}_{j}} = 
  \frac{r_{i}+r_{i}^{Mut}}{\alpha_{i}+c_{i}r_{i}^{Mut}} = 
  \frac{r_{i}^{*+}}{r_{i}^{*-}}  \nonumber\\
  \\
  N^{*}_{j}=\frac{r_{j}+\sum_{i=1}^{n_{p}}b_{ji}N^{*}_{i}}{\alpha_{j}+c_{j}\sum_{i=1}^{n_{a}}b_{ij}N^{*}_{i}} =
  \frac{r_{j}+r_{j}^{Mut}}{\alpha_{j}+c_{j}r_{j}^{Mut}} =
  \frac{r_{j}^{*+}}{r_{j}^{*-}}
  \end{array}
\end{eqnarray} 
 
\noindent where the rates $r^{*+}$ and $r^{*-}$ stand for the \emph{positive effective growth rate} and the \emph{per capita negative effective growth rate}, respectively.
 
The system [\ref{eq:N_especies}] can also be expanded around the singular point.

\small
\begin{equation}
\begin{array}{lcl}
\textstyle \frac{dN_{i}}{dt}=r_{i}+\sum\limits_{j=1}^{n_{a}}b_{ij}(N^{*}_{j}+\tilde{N}_{j})- (\alpha_{i}+c_{i}\sum\limits_{j=1}^{n_{a}}b_{ij}(N^{*}_j+\tilde{N}_{j}))(N^{*}_i+\tilde{N}_{i}) \nonumber\\
\textstyle \frac{dN_{j}}{dt}=r_{j}+\sum\limits_{i=1}^{n_{p}}b_{ji}(N^{*}_{i}+\tilde{N}_{i})-(\alpha_{j}+c_{j}\sum\limits_{i=1}^{n_{p}}b_{ji}(N^{*}_i+\tilde{N}_{i}))(N^{*}_j+\tilde{N}_{j}) 
\stepcounter{equation}\tag{\theequation}\label{eq:effrateTaylorN}
\end{array}
\end{equation}
\normalsize

\noindent where the subscript $i$ stands for plant species and the subscript $j$ stands for animal species. 

The set of $n_{a} + n_{p}$  equations can also be rewritten retaining only the linear terms as:
\small
\begin{align}
\begin{array}{lcl}
\displaystyle \frac{dN_{i}}{dt} = \sum_{j=1}^{n_{a}} \tilde{N}_{j} \left(  b_{ij} - c_{i} b_{ij}\, N^{*}_i\right) - \tilde{N}_{i}(\alpha_{i}+ c_{i} \sum_{j=1}^{n_{a}} b_{ij} \, N^{*}_{j})\nonumber\\
\displaystyle \frac{dN_{j}}{dt} = \sum_{i=1}^{n_{p}} \tilde{N}_{i} \left( b_{ji} - c_{j} b_{ji}\, N^{*}_j\right) - \tilde{N}_{j}(\alpha_{j}+ c_{j} \sum_{i=1}^{n_{p}} b_{ji} \, N^{*}_{i})\stepcounter{equation}\tag{\theequation}\label{eq:effrateTaylor2N}
\end{array}
\end{align}
\normalsize

The coefficients of $\tilde{N}_{i,j}$ are the entries of the Jacobian matrix. 
The absolute values of the diagonal terms (for any i-plant species and any j-animal species) are:

\begin{align}
\displaystyle & J_{ii}=N^{*}_{i}\left(\alpha_{i} + c_{i} \sum_{j=1}^{n_{a}} b_{ij} N^{*}_{j}\right) \nonumber\\
\displaystyle & J_{jj}=N^{*}_{j}\left(\alpha_{j} + c_{j} \sum_{i=1}^{n_{p}} b_{ji} N^{*}_{i}\right)
\label{eq:Jii2}
\end{align}

\noindent and the non-diagonal terms:

\begin{align}
\displaystyle & J_{ij}=N^{*}_{i}b_{ij}\left( 1-c_{i}N^{*}_{i}\right)\nonumber\\
\displaystyle & J_{ji}=N^{*}_{j}b_{ji}\left( 1-c_{j}N^{*}_{j}\right)
\label{eq:Jij}
\end{align}

So, the Jacobian matrix can be written as:

\begin{equation*}
J=\left(
   \begin{array}{ccccc}
      \ddots  & \cdots & \cdots & \cdots & \cdots \\
      \cdots  & -J_{ii} & \cdots & J_{ij} & \cdots \\
      \vdots  & \vdots & \ddots  & \vdots & \vdots  \\
      \cdots  & J_{ji} & \cdots & -J_{jj} & \cdots \\
      \cdots  & \cdots & \cdots & \cdots  & \ddots
   \end{array}
\right)
\end{equation*}

\noindent where the diagonal entries are all negatives and the off-diagonal terms are all positives.

The sum of eigenvalues satisfy:

\begin{equation}
  \sum_{k}^{n_{a}+n_{p}} \lambda_{k}= -\left(\sum_{k}^{n_{a}+n_{p}} J_{kk}\right)
  \stepcounter{equation}\tag{\theequation}\label{eq:sum_lambdas}
\end{equation}

This means that not all the eigenvalues are positives, and then the \emph{singular point} is not an asymptotically unstable node. On the other hand the eigenvalues cannot be complex because all the terms of the Jacobian matrix out of the diagonal are zero or positives values so they are stable nodes or saddle points.

\newpage

\section{Numerical treatment of the equations}
\label{NumSim}

Population models deal with sets of discrete entities such as animals or plants and computer simulation is a powerful tool to describe the dynamics and stochastic behavior. The choice of a specific simulation method depends on its accuracy and computational efficiency, and sometimes is a challenge.

For instance, Discrete Markov models have been frequently used for this kind of simulation, but this approach has a number of disadvantages compared with Discrete Stochastic Simulation (Poisson simulations or Binomial Simulations). In moderate size Markov models, the set of states may be huge, while Binomial or Poisson Simulation aggregate state variables make them much faster \citep{gustafsson2007bringing, balcan2009multiscale}.

We have chosen Binomial Simulation to solve the equations of our mutualistic population model. This technique is a stochastic extension of Continuous System Simulation and a reasonable choice when the outcome of the random process has only two values. For instance, survival over a finite time interval is a Bernoulli process, the individual either lives or dies. Breeding may also be described by a Bernouilli trial if time interval is small. 

For a species with intrinsic growth $r$, we can assume that probability of breeding over an interval $\Delta T$ is exponentially distributed with an average value $1/r$. So, the probability of reproduction is:
\begin{equation}
\label{eq:probbreeding}
P = \int_0^{\Delta T} \! e^{-r\, T}  \, dt = 1 - e^{-r\, \Delta T}
\end{equation}
In particular, a population of $N$ individuals in time $t$, with pure exponential growth, will be in $t+\Delta T$:
\begin{equation}
N(t+\Delta T)=N(t) + sgn \left(r \right) Binomial \left( N(t),P \right)
\end{equation}
The set of equations \eqref{eq:modeloralphaconmut} becomes in stochastic form:
\begin{equation}
\begin{split}
N^{a}_{j}(t+\Delta T)=N^{a}_{j}(t) + sgn \left(\hat{r}^{a}_{ef,j} \right) Binomial \left( N^{a}_{j}(t),P^{a}_{j}\right)\\
N^{p}_{l}(t+\Delta T)=N^{p}_{l}(t) + sgn \left(\hat{r}^{p}_{ef,l} \right) Binomial \left(N^{p}_{l}(t),P^{p}_{l} \right)
\end{split}
\end{equation}
where $\hat{r}^{a}_{ef,j}$ is the class \emph{a} $j$th-species effective growth rate in the simulation period, and $P^{a}_{j}, P^{p}_{l}$ , the probabilities of growth according to equation \ref{eq:probbreeding}. In particular, working with one day steps, as we do:
\begin{equation}
\hat{r}_{ef} = e^{r_{ef}/365}-1
\end{equation}

\newpage

\section{Data tables}
\label{DataTables}
\begin{table}[h!]
\label{exp1}
\centering
\footnotesize
\begin{tabular}{lrrrr}
\hline
 & Pl 1 & Pl 2 & Pl 3 & Pl 4  \\
\hline
$b_{1j}${\tiny $\left(10^{-6}\right)$} & 1 & 12 & 12 & 16\\
$b_{2j}${\tiny $\left(10^{-6}\right)$} & 12 & 4 & 11 & 0 \\
$b_{3j}${\tiny $\left(10^{-6}\right)$} & 12 & 10 & 0 & 0 \\
$b_{4j}${\tiny $\left(10^{-6}\right)$} & 6 & 10 & 0 & 0 \\
$b_{5j}${\tiny $\left(10^{-6}\right)$} & 10 & 0 & 0 & 0 \\
$N_{init\,j}$ & 700 & 600 & 500 & 200 \\
$c_{j}${\tiny $\left(10^{-4}\right)$} & 1 & 1 & 1 & 1 \\
$\alpha_{j}${\tiny $\left(10^{-6}\right)$} & 7 & 12 & 12 & 10 \\
$r_{birth\, j}$ & 0.004 & 0.01 & 0.01 & 0.005 \\
$r_{death\, j}$ & 0.005 & 0.04 & 0.05 & 0.0055 \\
\hline
\\
\end{tabular}
\begin{tabular}{lrrrrr}
\hline
 &Pol 1&Pol 2&Pol 3&Pol 4&Pol 5\\
\hline
$b_{1m}${\tiny $\left(10^{-6}\right)$}&14&13&10&10&20\\
$b_{2m}${\tiny $\left(10^{-6}\right)$}&12&6&1&10&0\\
$b_{3m}${\tiny $\left(10^{-6}\right)$}&2&5&1&0&0\\
$b_{4m}${\tiny $\left(10^{-6}\right)$}&10&1&0&0&0\\
$N_{init\,m}$ & 500 & 300 & 500 & 200 & 150 \\
$c_{m}${\tiny $\left(10^{-4}\right)$} & 1 & 1 & 1 & 1 & 1\\
$\alpha_{m}${\tiny $\left(10^{-6}\right)$} & 10 & 10 & 8 & 10 & 30\\
$r_{b\, m}$ & 0.28 & 0.02 & 0.05 & 0.02 & 0.02 \\
$r_{d\, m}$ & 0.44 & 0.058 & 0.065 & 0.034 & 0.038 \\
\hline
\end{tabular}
\normalsize
\caption{Mutualistic coefficients and conditions for the first simulation (fig. \ref{fig:red_exper_stab1}). Top, pollinator-plant interaction matrix; bottom, plant-pollinator matrix}
\label{tab:experiment1}
\end{table}
\begin{table}[h!]
\label{exp2}
\centering
\footnotesize
\begin{tabular}{lrrrr}
\hline
 & Pl 1 & Pl 2 & Pl 3 & Pl 4  \\
\hline
$b_{1j}${\tiny $\left(10^{-6}\right)$} & 1 & 12 & 12 & 16\\
$b_{2j}${\tiny $\left(10^{-6}\right)$} & 12 & 4 & 11 & 0 \\
$b_{3j}${\tiny $\left(10^{-6}\right)$} & 12 & 10 & 0 & 0 \\
$b_{4j}${\tiny $\left(10^{-6}\right)$} & 6 & 10 & 0 & 0 \\
$b_{5j}${\tiny $\left(10^{-6}\right)$} & 10 & 0 & 0 & 0 \\
$N_{init\,j}$ & 1500 & 2000 & 1200 & 1500 \\
$c_{j}${\tiny $\left(10^{-4}\right)$} & 1 & 1 & 1 & 1 \\
$\alpha_{j}${\tiny $\left(10^{-6}\right)$} & 7 & 12 & 12 & 10 \\
$r_{birth\, j}$ & 0.004 & 0.01 & 0.01 & 0.005 \\
$r_{death\, j}$ & 0.005 & 0.04 & 0.05 & 0.0055 \\
\hline
\\
\end{tabular}
\centering
\begin{tabular}{lrrrrr}
\hline
 &Pol 1&Pol 2&Pol 3&Pol 4&Pol 5\\
\hline
$b_{1m}${\tiny $\left(10^{-6}\right)$}&14&13&10&10&20\\
$b_{2m}${\tiny $\left(10^{-6}\right)$}&12&6&1&10&0\\
$b_{3m}${\tiny $\left(10^{-6}\right)$}&2&5&1&0&0\\
$b_{4m}${\tiny $\left(10^{-6}\right)$}&10&1&0&0&0\\
$N_{init\,m}$ & 700 & 600 & 1000 & 700 & 500 \\
$c_{m}${\tiny $\left(10^{-4}\right)$} & 1 & 1 & 1 & 1 & 1\\
$\alpha_{m}${\tiny $\left(10^{-6}\right)$} & 10 & 10 & 8 & 10 & 30\\
$r_{b\, m}$ & 0.28 & 0.02 & 0.05 & 0.02 & 0.02 \\
$r_{d\, m}$ & 0.44 & 0.058 & 0.065 & 0.034 & 0.038 \\
\hline
\end{tabular}
\normalsize
\caption{Mutualistic coefficients and conditions for the second simulation (fig. \ref{fig:exper_stab2}).}
\label{tab:experiment2}
\end{table}

\begin{table}[h!]
\scriptsize
\begin{tabular}{lrrrrrrr}
\hline
 &Pl 1&Pl 2&Pl 3&Pl 4&Pl 5&Pl 6&Pl 7\\
\hline
$b_{1j\, }${\tiny $\left(10^{-6}\right)$}&20&12&16&16&19&25&35\\
$b_{2j\, }${\tiny $\left(10^{-6}\right)$}&12&14&4.1&2&22&0&0\\
$b_{3j\, }${\tiny $\left(10^{-6}\right)$}&20&11&3.1&20&0&0&0\\
$b_{4j\, }${\tiny $\left(10^{-6}\right)$}&11&24&0&0&0&0&0\\
$b_{5j\, }${\tiny $\left(10^{-6}\right)$}&1&0&0&0&0&0&0\\
$N_{init\,j}$&1200 & 1500 & 800 & 770 & 700 & 800 & 400\\
$c_{j}${\tiny $\left(10^{-4}\right)$} & 1 & 0.5 & 1 & 2 & 1 & 1 & 1\\
$\alpha_{j}${\tiny $\left(10^{-6}\right)$} & 20 & 30 & 10 & 10 & 50 & 10 &10\\
$r_{birth\, j}$ & 0.004 & 0.01 & 0.02 & 0.005 & 0.004 & 0.02 & 0.025\\
$r_{death\, j}$ & 0.03 & 0.04 & 0.04 & 0.055 & 0.03 & 0.03 & 0.028\\
\hline
\\
\end{tabular}
\centering
\begin{tabular}{lrrrrr}
\hline
 &Pol 1&Pol 2&Pol 3&Pol 4&Pol 5\\
\hline
$b_{1m\,}${\tiny $\left(10^{-6}\right)$}&14&13&23&30&23\\
$b_{2m\,}${\tiny $\left(10^{-6}\right)$}&19&26&10&10&0\\
$b_{3m\,}${\tiny $\left(10^{-6}\right)$}&2&25&10&0&0\\
$b_{4m\,}${\tiny $\left(10^{-6}\right)$}&1&11&10&0&0\\
$b_{5m\,}${\tiny $\left(10^{-6}\right)$}&1&1&0&0&0\\
$b_{6m}${\tiny $\left(10^{-6}\right)$}&1&0&0&0&0\\
$b_{7m}${\tiny $\left(10^{-6}\right)$}&1&0&0&0&0\\
$N_{init\,m}$ & 1200 & 1500 & 1300 & 1000 & 700 \\
$c_{m}${\tiny $\left(10^{-4}\right)$} & 1 & 1 & 1 & 0.7 & 2\\
$\alpha_{m}${\tiny $\left(10^{-6}\right)$} & 10 & 10 & 20 & 10 & 20\\
$r_{b\, m}$ & 0.08 & 0.02 & 0.02 & 0.05 & 0.02 \\
$r_{d\, m}$ & 0.11 & 0.078 & 0.068 & 0.07 & 0.028 \\
\hline
\end{tabular}
\normalsize
\caption{Mutualistic coefficients and conditions for the simulation of a high nested network (fig. \ref{fig:exper_resilience_strong}).}
\label{tab:exper_resilience_strong}
\end{table}
\begin{table}[h!]
\centering
\scriptsize
\begin{tabular}{lrrrrrrr}
\hline
 &Pl 1&Pl 2&Pl 3&Pl 4&Pl 5&Pl 6&Pl 7\\
\hline
$b_{1j\, }${\tiny $\left(10^{-6}\right)$}&20&12&16&16&19&0&45\\
$b_{2j\, }${\tiny $\left(10^{-6}\right)$}&12&14&4.1&2&22&0&0\\
$b_{3j\, }${\tiny $\left(10^{-6}\right)$}&20&11&3.1&20&0&0&0\\
$b_{4j\, }${\tiny $\left(10^{-6}\right)$}&11&24&0&0&0&0&0\\
$b_{5j\, }${\tiny $\left(10^{-6}\right)$}&0&0&0&0&0&25&1\\
$N_{init\,j}$&1200 & 1500 & 800 & 770 & 700 & 400 &1000\\
$c_{j}${\tiny $\left(10^{-4}\right)$} & 1 & 0.5 & 1 & 2 & 1 & 1 & 1\\
$\alpha_{j}${\tiny $\left(10^{-6}\right)$} & 20 & 30 & 10 & 10 & 50 & 10 &10\\
$r_{birth\, j}$ & 0.004 & 0.01 & 0.02 & 0.005 & 0.004 & 0.02 & 0.025\\
$r_{death\, j}$ & 0.03 & 0.04 & 0.04 & 0.055 & 0.03 & 0.024 & 0.04\\
\hline
\\
\end{tabular}
\begin{tabular}{lrrrrr}
\hline
 &Pol 1&Pol 2&Pol 3&Pol 4&Pol 5\\
\hline
$b_{1m\,}${\tiny $\left(10^{-6}\right)$}&14&13&23&30&0\\
$b_{2m\,}${\tiny $\left(10^{-6}\right)$}&19&26&10&10&0\\
$b_{3m\,}${\tiny $\left(10^{-6}\right)$}&2&25&10&0&0\\
$b_{4m\,}${\tiny $\left(10^{-6}\right)$}&1&11&10&0&0\\
$b_{5m\,}${\tiny $\left(10^{-6}\right)$}&1&1&0&0&0\\
$b_{6m}${\tiny $\left(10^{-6}\right)$}&0&0&0&0&5\\
$b_{7m}${\tiny $\left(10^{-6}\right)$}&1&0&0&0&30\\
$N_{init\,m}$ & 1200 & 1500 & 1300 & 1000 & 700 \\
$c_{m}${\tiny $\left(10^{-4}\right)$} & 1 & 1 & 1 & 0.7 & 2\\
$\alpha_{m}${\tiny $\left(10^{-6}\right)$} & 10 & 10 & 20 & 10 & 20\\
$r_{b\, m}$ & 0.09 & 0.02 & 0.02 & 0.05 & 0.02 \\
$r_{d\, m}$ & 0.11 & 0.058 & 0.04 & 0.07 & 0.025 \\
\hline
\end{tabular}
\normalsize
\caption{Mutualistic coefficients and conditions for the simulation of low nested network (fig. \ref{fig:red_exper_resilience_weak}).}
\label{tab:exper_resilience_weak}
\end{table}

\newpage
\bibliography{ref-mutualism}

\begin{thebibliography}{38}
\expandafter\ifx\csname natexlab\endcsname\relax\def\natexlab#1{#1}\fi
\expandafter\ifx\csname bibnamefont\endcsname\relax
  \def\bibnamefont#1{#1}\fi
\expandafter\ifx\csname bibfnamefont\endcsname\relax
  \def\bibfnamefont#1{#1}\fi
\expandafter\ifx\csname citenamefont\endcsname\relax
  \def\citenamefont#1{#1}\fi
\expandafter\ifx\csname url\endcsname\relax
  \def\url#1{\texttt{#1}}\fi
\expandafter\ifx\csname urlprefix\endcsname\relax\def\urlprefix{URL }\fi
\providecommand{\bibinfo}[2]{#2}
\providecommand{\eprint}[2][]{\url{#2}}

\bibitem[{\citenamefont{Kennedy and Norman}(2005)}]{Kennedy05}
\bibinfo{author}{\bibfnamefont{D.}~\bibnamefont{Kennedy}} \bibnamefont{and}
  \bibinfo{author}{\bibfnamefont{C.}~\bibnamefont{Norman}},
  \bibinfo{journal}{Science} \textbf{\bibinfo{volume}{309}},
  \bibinfo{pages}{75} (\bibinfo{year}{2005}).

\bibitem[{\citenamefont{Pennisi}(2005)}]{Pennisi05}
\bibinfo{author}{\bibfnamefont{E.}~\bibnamefont{Pennisi}},
  \bibinfo{journal}{Science} \textbf{\bibinfo{volume}{309}},
  \bibinfo{pages}{90} (\bibinfo{year}{2005}).

\bibitem[{\citenamefont{Stokstad}(2005)}]{Stokstad05}
\bibinfo{author}{\bibfnamefont{E.}~\bibnamefont{Stokstad}},
  \bibinfo{journal}{Science} \textbf{\bibinfo{volume}{309}},
  \bibinfo{pages}{102} (\bibinfo{year}{2005}).

\bibitem[{\citenamefont{Williams and Mart\'{\i}nez}(2000)}]{williams00}
\bibinfo{author}{\bibfnamefont{R.}~\bibnamefont{Williams}} \bibnamefont{and}
  \bibinfo{author}{\bibfnamefont{N.~D.} \bibnamefont{Mart\'{\i}nez}},
  \bibinfo{journal}{Nature} \textbf{\bibinfo{volume}{404}},
  \bibinfo{pages}{180} (\bibinfo{year}{2000}).

\bibitem[{\citenamefont{Dunne et~al.}(2002)\citenamefont{Dunne, Williams, and
  Mart\'{\i}nez}}]{dunne02}
\bibinfo{author}{\bibfnamefont{J.~A.} \bibnamefont{Dunne}},
  \bibinfo{author}{\bibfnamefont{R.~J.} \bibnamefont{Williams}},
  \bibnamefont{and} \bibinfo{author}{\bibfnamefont{N.~D.}
  \bibnamefont{Mart\'{\i}nez}}, \bibinfo{journal}{Ecology Letters}
  \textbf{\bibinfo{volume}{5}}, \bibinfo{pages}{558} (\bibinfo{year}{2002}).

\bibitem[{\citenamefont{Olesen et~al.}(2007)\citenamefont{Olesen, Bascompte,
  Dupont, and Jordano}}]{olensen07}
\bibinfo{author}{\bibfnamefont{J.~M.} \bibnamefont{Olesen}},
  \bibinfo{author}{\bibfnamefont{J.}~\bibnamefont{Bascompte}},
  \bibinfo{author}{\bibfnamefont{Y.~L.} \bibnamefont{Dupont}},
  \bibnamefont{and} \bibinfo{author}{\bibfnamefont{P.}~\bibnamefont{Jordano}},
  \bibinfo{journal}{Proceedings of the National Academy of Sciences USA}
  \textbf{\bibinfo{volume}{104}}, \bibinfo{pages}{19891}
  (\bibinfo{year}{2007}).

\bibitem[{\citenamefont{Allesina et~al.}(2008)\citenamefont{Allesina, Alonso,
  and Pascual}}]{allesina08}
\bibinfo{author}{\bibfnamefont{S.}~\bibnamefont{Allesina}},
  \bibinfo{author}{\bibfnamefont{D.}~\bibnamefont{Alonso}}, \bibnamefont{and}
  \bibinfo{author}{\bibfnamefont{M.}~\bibnamefont{Pascual}},
  \bibinfo{journal}{Science} \textbf{\bibinfo{volume}{320}},
  \bibinfo{pages}{658} (\bibinfo{year}{2008}).

\bibitem[{\citenamefont{Bascompte}(2009)}]{bascompte09}
\bibinfo{author}{\bibfnamefont{J.}~\bibnamefont{Bascompte}},
  \bibinfo{journal}{Science} \textbf{\bibinfo{volume}{325}},
  \bibinfo{pages}{416} (\bibinfo{year}{2009}).

\bibitem[{\citenamefont{Saavedra et~al.}(2009)\citenamefont{Saavedra,
  Reed-Tsochas, and Uzzi}}]{saavedra09}
\bibinfo{author}{\bibfnamefont{S.}~\bibnamefont{Saavedra}},
  \bibinfo{author}{\bibfnamefont{F.}~\bibnamefont{Reed-Tsochas}},
  \bibnamefont{and} \bibinfo{author}{\bibfnamefont{B.}~\bibnamefont{Uzzi}},
  \bibinfo{journal}{Nature} \textbf{\bibinfo{volume}{457}},
  \bibinfo{pages}{463} (\bibinfo{year}{2009}).

\bibitem[{\citenamefont{Bastolla et~al.}(2009)\citenamefont{Bastolla, Fortuna,
  Alberto Pascual-Garc\'{\i}a, Ferrera, Luque, and Bascompte}}]{bastolla09}
\bibinfo{author}{\bibfnamefont{U.}~\bibnamefont{Bastolla}},
  \bibinfo{author}{\bibfnamefont{M.~A.} \bibnamefont{Fortuna}},
  \bibinfo{author}{\bibfnamefont{A.}~\bibnamefont{Alberto
  Pascual-Garc\'{\i}a}},
  \bibinfo{author}{\bibfnamefont{A.}~\bibnamefont{Ferrera}},
  \bibinfo{author}{\bibfnamefont{B.}~\bibnamefont{Luque}}, \bibnamefont{and}
  \bibinfo{author}{\bibfnamefont{J.}~\bibnamefont{Bascompte}},
  \bibinfo{journal}{Nature} \textbf{\bibinfo{volume}{458}},
  \bibinfo{pages}{1018} (\bibinfo{year}{2009}).

\bibitem[{\citenamefont{Fortuna et~al.}(2010)\citenamefont{Fortuna, Stouffer,
  Olesen, Jordano, Mouillot, Krasnov, Poulin, and
  Bascompte}}]{fortuna2010nestedness}
\bibinfo{author}{\bibfnamefont{M.~A.} \bibnamefont{Fortuna}},
  \bibinfo{author}{\bibfnamefont{D.~B.} \bibnamefont{Stouffer}},
  \bibinfo{author}{\bibfnamefont{J.~M.} \bibnamefont{Olesen}},
  \bibinfo{author}{\bibfnamefont{P.}~\bibnamefont{Jordano}},
  \bibinfo{author}{\bibfnamefont{D.}~\bibnamefont{Mouillot}},
  \bibinfo{author}{\bibfnamefont{B.~R.} \bibnamefont{Krasnov}},
  \bibinfo{author}{\bibfnamefont{R.}~\bibnamefont{Poulin}}, \bibnamefont{and}
  \bibinfo{author}{\bibfnamefont{J.}~\bibnamefont{Bascompte}},
  \bibinfo{journal}{Journal of Animal Ecology} \textbf{\bibinfo{volume}{79}},
  \bibinfo{pages}{811} (\bibinfo{year}{2010}).

\bibitem[{\citenamefont{Encinas-Viso et~al.}(2012)\citenamefont{Encinas-Viso,
  Revilla, and Etienne}}]{encinas12}
\bibinfo{author}{\bibfnamefont{F.}~\bibnamefont{Encinas-Viso}},
  \bibinfo{author}{\bibfnamefont{T.~A.} \bibnamefont{Revilla}},
  \bibnamefont{and} \bibinfo{author}{\bibfnamefont{R.~S.}
  \bibnamefont{Etienne}}, \bibinfo{journal}{Ecology Letters}
  \textbf{\bibinfo{volume}{15}}, \bibinfo{pages}{198} (\bibinfo{year}{2012}).

\bibitem[{\citenamefont{Sigler}(2002)}]{Sigler02}
\bibinfo{author}{\bibfnamefont{L.~E. L.~E.} \bibnamefont{Sigler}},
  \emph{\bibinfo{title}{{Fibonacci}'s Liber Abaci: {A} Translation into Modern
  {English} of {Leonardo Pisano}'s {Book of Calculation}}}, Sources and studies
  in the history of mathematics and physical sciences
  (\bibinfo{publisher}{Springer}, \bibinfo{year}{2002}).

\bibitem[{\citenamefont{Malthus}(1798)}]{Malthus98}
\bibinfo{author}{\bibfnamefont{T.~R.} \bibnamefont{Malthus}},
  \emph{\bibinfo{title}{An essay on the principle of population, or, A view of
  its past and present effects on human happiness [electronic resource] : with
  an inquiry into our prospects respecting the future removal on mitigation of
  the evils which it occasions / by T.R. Malthus}} (\bibinfo{publisher}{Roger
  Chew Weightman, Washington}, \bibinfo{year}{1798}), \bibinfo{edition}{1st}
  ed.

\bibitem[{\citenamefont{Verhulst}(1845)}]{Verhulst1845}
\bibinfo{author}{\bibfnamefont{P.~F.} \bibnamefont{Verhulst}},
  \bibinfo{journal}{Nouveaux Memoires de l'Academie Royale des Sciences et
  Belles-Lettres de Bruxelles} \textbf{\bibinfo{volume}{18}},
  \bibinfo{pages}{1} (\bibinfo{year}{1845}).

\bibitem[{\citenamefont{Mallet}(2012)}]{mallet2012struggle}
\bibinfo{author}{\bibfnamefont{J.}~\bibnamefont{Mallet}},
  \bibinfo{journal}{Evolutionary Ecology Research}
  \textbf{\bibinfo{volume}{14}}, \bibinfo{pages}{627} (\bibinfo{year}{2012}).

\bibitem[{\citenamefont{Kuno}(1991)}]{kuno1991some}
\bibinfo{author}{\bibfnamefont{E.}~\bibnamefont{Kuno}},
  \bibinfo{journal}{Researches on population ecology}
  \textbf{\bibinfo{volume}{33}}, \bibinfo{pages}{33} (\bibinfo{year}{1991}).

\bibitem[{\citenamefont{Gabriel et~al.}(2005)\citenamefont{Gabriel, Saucy, and
  Bersier}}]{gabriel2005paradoxes}
\bibinfo{author}{\bibfnamefont{J.-P.} \bibnamefont{Gabriel}},
  \bibinfo{author}{\bibfnamefont{F.}~\bibnamefont{Saucy}}, \bibnamefont{and}
  \bibinfo{author}{\bibfnamefont{L.-F.} \bibnamefont{Bersier}},
  \bibinfo{journal}{Ecological Modelling} \textbf{\bibinfo{volume}{185}},
  \bibinfo{pages}{147} (\bibinfo{year}{2005}).

\bibitem[{\citenamefont{Volterra}(1926)}]{Volterra26}
\bibinfo{author}{\bibfnamefont{V.}~\bibnamefont{Volterra}},
  \bibinfo{journal}{Nature} \textbf{\bibinfo{volume}{118}},
  \bibinfo{pages}{558} (\bibinfo{year}{1926}).

\bibitem[{\citenamefont{Darwin}(1862)}]{Darwin62}
\bibinfo{author}{\bibfnamefont{C.}~\bibnamefont{Darwin}},
  \emph{\bibinfo{title}{On the Various Contrivances by Which British and
  Foreign Orchids are fertilised by Insects, and on the Good Effects of
  Intercrossing}} (\bibinfo{publisher}{Murray, London}, \bibinfo{year}{1862}).

\bibitem[{\citenamefont{Ehrlich and Raven}(1964)}]{Ehr64}
\bibinfo{author}{\bibfnamefont{P.}~\bibnamefont{Ehrlich}} \bibnamefont{and}
  \bibinfo{author}{\bibfnamefont{P.}~\bibnamefont{Raven}},
  \bibinfo{journal}{Evolution} \textbf{\bibinfo{volume}{18}},
  \bibinfo{pages}{586} (\bibinfo{year}{1964}).

\bibitem[{\citenamefont{May}(1981)}]{may1981models}
\bibinfo{author}{\bibfnamefont{R.}~\bibnamefont{May}},
  \emph{\bibinfo{title}{Models for two interacting populations. in theoretical
  ecology. principles and applications. 2nd edn.(ed. rm may.) pp. 78--104}}
  (\bibinfo{year}{1981}).

\bibitem[{\citenamefont{Wright}(1989)}]{Wright89}
\bibinfo{author}{\bibfnamefont{D.~H.} \bibnamefont{Wright}},
  \bibinfo{journal}{The American Naturalist} \textbf{\bibinfo{volume}{134}},
  \bibinfo{pages}{664} (\bibinfo{year}{1989}).

\bibitem[{\citenamefont{Bastolla et~al.}(2005)\citenamefont{Bastolla,
  L{\"a}ssig, Manrubia, and Valleriani}}]{Bastolla05}
\bibinfo{author}{\bibfnamefont{U.}~\bibnamefont{Bastolla}},
  \bibinfo{author}{\bibfnamefont{M.}~\bibnamefont{L{\"a}ssig}},
  \bibinfo{author}{\bibfnamefont{S.~C.} \bibnamefont{Manrubia}},
  \bibnamefont{and}
  \bibinfo{author}{\bibfnamefont{A.}~\bibnamefont{Valleriani}},
  \bibinfo{journal}{J Theor Biol} \textbf{\bibinfo{volume}{235}},
  \bibinfo{pages}{531} (\bibinfo{year}{2005}).

\bibitem[{\citenamefont{Johnson and Amarasekare}(2013)}]{johnson13}
\bibinfo{author}{\bibfnamefont{C.~A.} \bibnamefont{Johnson}} \bibnamefont{and}
  \bibinfo{author}{\bibfnamefont{P.}~\bibnamefont{Amarasekare}},
  \bibinfo{journal}{Journal of Theoretical Biology}
  \textbf{\bibinfo{volume}{328}}, \bibinfo{pages}{54} (\bibinfo{year}{2013}).

\bibitem[{\citenamefont{Th{\'e}bault and
  Fontaine}(2010)}]{thebault2010stability}
\bibinfo{author}{\bibfnamefont{E.}~\bibnamefont{Th{\'e}bault}}
  \bibnamefont{and} \bibinfo{author}{\bibfnamefont{C.}~\bibnamefont{Fontaine}},
  \bibinfo{journal}{Science} \textbf{\bibinfo{volume}{329}},
  \bibinfo{pages}{853} (\bibinfo{year}{2010}).

\bibitem[{\citenamefont{Staniczenko et~al.}(2013)\citenamefont{Staniczenko,
  Kopp, and Allesina}}]{staniczenko2013ghost}
\bibinfo{author}{\bibfnamefont{P.~P.} \bibnamefont{Staniczenko}},
  \bibinfo{author}{\bibfnamefont{J.~C.} \bibnamefont{Kopp}}, \bibnamefont{and}
  \bibinfo{author}{\bibfnamefont{S.}~\bibnamefont{Allesina}},
  \bibinfo{journal}{Nature communications} \textbf{\bibinfo{volume}{4}},
  \bibinfo{pages}{1391} (\bibinfo{year}{2013}).

\bibitem[{\citenamefont{Stenseth et~al.}(1998)\citenamefont{Stenseth, Falck,
  Chan, Bj{\o}rnstad, O'Donoghue, Tong, Boonstra, Boutin, Krebs, and
  Yoccoz}}]{stenseth98}
\bibinfo{author}{\bibfnamefont{N.~C.} \bibnamefont{Stenseth}},
  \bibinfo{author}{\bibfnamefont{W.}~\bibnamefont{Falck}},
  \bibinfo{author}{\bibfnamefont{K.-S.} \bibnamefont{Chan}},
  \bibinfo{author}{\bibfnamefont{O.~N.} \bibnamefont{Bj{\o}rnstad}},
  \bibinfo{author}{\bibfnamefont{M.}~\bibnamefont{O'Donoghue}},
  \bibinfo{author}{\bibfnamefont{H.}~\bibnamefont{Tong}},
  \bibinfo{author}{\bibfnamefont{R.}~\bibnamefont{Boonstra}},
  \bibinfo{author}{\bibfnamefont{S.}~\bibnamefont{Boutin}},
  \bibinfo{author}{\bibfnamefont{C.~J.} \bibnamefont{Krebs}}, \bibnamefont{and}
  \bibinfo{author}{\bibfnamefont{N.~G.} \bibnamefont{Yoccoz}},
  \bibinfo{journal}{Proceedings of the National Academy of Sciences USA}
  \textbf{\bibinfo{volume}{95}}, \bibinfo{pages}{15430} (\bibinfo{year}{1998}).

\bibitem[{\citenamefont{Krebs}(2002)}]{krebs02}
\bibinfo{author}{\bibfnamefont{C.~J.} \bibnamefont{Krebs}},
  \bibinfo{journal}{Proc. R. Soc. Lond. B} \textbf{\bibinfo{volume}{357}},
  \bibinfo{pages}{1211} (\bibinfo{year}{2002}).

\bibitem[{\citenamefont{Rueness et~al.}(2003)\citenamefont{Rueness, Stenseth,
  O'Donoghue, Boutin, Ellegren, and Jakobsen}}]{rueness03}
\bibinfo{author}{\bibfnamefont{E.~K.} \bibnamefont{Rueness}},
  \bibinfo{author}{\bibfnamefont{N.~C.} \bibnamefont{Stenseth}},
  \bibinfo{author}{\bibfnamefont{M.}~\bibnamefont{O'Donoghue}},
  \bibinfo{author}{\bibfnamefont{S.}~\bibnamefont{Boutin}},
  \bibinfo{author}{\bibfnamefont{H.}~\bibnamefont{Ellegren}}, \bibnamefont{and}
  \bibinfo{author}{\bibfnamefont{K.~S.} \bibnamefont{Jakobsen}},
  \bibinfo{journal}{Nature} \textbf{\bibinfo{volume}{425}}, \bibinfo{pages}{69}
  (\bibinfo{year}{2003}).

\bibitem[{\citenamefont{Tyler et~al.}(2008)\citenamefont{Tyler, Forchhammer,
  and {\O}ritsland}}]{tyler08}
\bibinfo{author}{\bibfnamefont{N.~J.~C.} \bibnamefont{Tyler}},
  \bibinfo{author}{\bibfnamefont{M.~C.} \bibnamefont{Forchhammer}},
  \bibnamefont{and} \bibinfo{author}{\bibfnamefont{N.~A.}
  \bibnamefont{{\O}ritsland}}, \bibinfo{journal}{Ecology}
  \textbf{\bibinfo{volume}{89}}, \bibinfo{pages}{1675} (\bibinfo{year}{2008}).

\bibitem[{\citenamefont{Jones et~al.}(2008)\citenamefont{Jones, Cockburn,
  Hamede, Hawkins, Hesterman, Lachish, Mann, McCallum, and
  Pemberton}}]{jones08}
\bibinfo{author}{\bibfnamefont{M.~E.} \bibnamefont{Jones}},
  \bibinfo{author}{\bibfnamefont{A.}~\bibnamefont{Cockburn}},
  \bibinfo{author}{\bibfnamefont{R.}~\bibnamefont{Hamede}},
  \bibinfo{author}{\bibfnamefont{C.}~\bibnamefont{Hawkins}},
  \bibinfo{author}{\bibfnamefont{H.}~\bibnamefont{Hesterman}},
  \bibinfo{author}{\bibfnamefont{S.}~\bibnamefont{Lachish}},
  \bibinfo{author}{\bibfnamefont{D.}~\bibnamefont{Mann}},
  \bibinfo{author}{\bibfnamefont{H.}~\bibnamefont{McCallum}}, \bibnamefont{and}
  \bibinfo{author}{\bibfnamefont{D.}~\bibnamefont{Pemberton}},
  \bibinfo{journal}{Proceedings of the National Academy of Sciences USA}
  \textbf{\bibinfo{volume}{105}}, \bibinfo{pages}{10023}
  (\bibinfo{year}{2008}).

\bibitem[{\citenamefont{Goh}(1979)}]{goh79}
\bibinfo{author}{\bibfnamefont{B.}~\bibnamefont{Goh}}, \bibinfo{journal}{The
  American Naturalist} \textbf{\bibinfo{volume}{113}}, \bibinfo{pages}{261}
  (\bibinfo{year}{1979}).

\bibitem[{\citenamefont{Suweis et~al.}(2013)\citenamefont{Suweis, Simini,
  Banavar, and Maritan}}]{suweis13}
\bibinfo{author}{\bibfnamefont{S.}~\bibnamefont{Suweis}},
  \bibinfo{author}{\bibfnamefont{F.}~\bibnamefont{Simini}},
  \bibinfo{author}{\bibfnamefont{J.}~\bibnamefont{Banavar}}, \bibnamefont{and}
  \bibinfo{author}{\bibfnamefont{A.}~\bibnamefont{Maritan}},
  \bibinfo{journal}{Nature} \textbf{\bibinfo{volume}{500}},
  \bibinfo{pages}{449} (\bibinfo{year}{2013}).

\bibitem[{\citenamefont{Balcan et~al.}(2009)\citenamefont{Balcan, Colizza,
  Gon{\c{c}}alves, Hu, Ramasco, and Vespignani}}]{balcan2009multiscale}
\bibinfo{author}{\bibfnamefont{D.}~\bibnamefont{Balcan}},
  \bibinfo{author}{\bibfnamefont{V.}~\bibnamefont{Colizza}},
  \bibinfo{author}{\bibfnamefont{B.}~\bibnamefont{Gon{\c{c}}alves}},
  \bibinfo{author}{\bibfnamefont{H.}~\bibnamefont{Hu}},
  \bibinfo{author}{\bibfnamefont{J.~J.} \bibnamefont{Ramasco}},
  \bibnamefont{and}
  \bibinfo{author}{\bibfnamefont{A.}~\bibnamefont{Vespignani}},
  \bibinfo{journal}{Proceedings of the National Academy of Sciences USA}
  \textbf{\bibinfo{volume}{106}}, \bibinfo{pages}{21484}
  (\bibinfo{year}{2009}).

\bibitem[{\citenamefont{Bascompte et~al.}(2003)\citenamefont{Bascompte,
  Jordano, Meli{\'a}n, and Olesen}}]{bascompte2003nested}
\bibinfo{author}{\bibfnamefont{J.}~\bibnamefont{Bascompte}},
  \bibinfo{author}{\bibfnamefont{P.}~\bibnamefont{Jordano}},
  \bibinfo{author}{\bibfnamefont{C.~J.} \bibnamefont{Meli{\'a}n}},
  \bibnamefont{and} \bibinfo{author}{\bibfnamefont{J.~M.}
  \bibnamefont{Olesen}}, \bibinfo{journal}{Proceedings of the National Academy
  of Sciences} \textbf{\bibinfo{volume}{100}}, \bibinfo{pages}{9383}
  (\bibinfo{year}{2003}).

\bibitem[{\citenamefont{Murray}(1993)}]{murraymathematical}
\bibinfo{author}{\bibfnamefont{J.}~\bibnamefont{Murray}},
  \emph{\bibinfo{title}{Mathematical biology i. an introduction.
  interdisciplinary applied mathematics 1993}} (\bibinfo{year}{1993}).

\bibitem[{\citenamefont{Gustafsson and Sternad}(2007)}]{gustafsson2007bringing}
\bibinfo{author}{\bibfnamefont{L.}~\bibnamefont{Gustafsson}} \bibnamefont{and}
  \bibinfo{author}{\bibfnamefont{M.}~\bibnamefont{Sternad}},
  \bibinfo{journal}{Mathematical biosciences} \textbf{\bibinfo{volume}{209}},
  \bibinfo{pages}{361} (\bibinfo{year}{2007}).

\end{thebibliography}

\end{document}